\documentclass[showpacs,preprintnumbers,amsmath,amssymb]{revtex4}

\usepackage{graphicx}
\usepackage{dcolumn}
\usepackage{bm}

\begin{document}

\preprint{APS/123-QED}

\title{Amplification
in the auditory periphery: the effect of coupling tuning mechanisms}

\author{K.A. Montgomery} 
\affiliation{ Mathematics Department,
University of Utah, Salt Lake City, UT 84112 }%

\author{M. Silber}
\affiliation{ Department of Engineering Sciences \& Applied
Mathematics, Northwestern University, Evanston, IL 60208 }%

\author{S.A. Solla}
\affiliation{Departments of Physics \& Astronomy and Physiology, 
Northwestern University, Evanston, IL 60208
}%

\date{\today}%

\begin{abstract}
A mathematical model describing the coupling between two independent
amplification mechanisms in auditory hair cells is proposed and
analyzed. Hair cells are cells in the inner ear responsible for
translating sound-induced mechanical stimuli into an electrical signal
that can then be recorded by the auditory nerve. In nonmammals, two
separate mechanisms have been postulated to contribute to the
amplification and tuning properties of the hair cells. Models of each
of these mechanisms have been shown to be poised near a Hopf
bifurcation. Through a weakly nonlinear analysis that assumes weak
periodic forcing, weak damping, and weak coupling, the
physiologically-based models of the two mechanisms are reduced to a
system of two coupled amplitude equations describing the resonant
response. The predictions that follow from an analysis of the reduced
equations, as well as performance benefits due to the coupling of the
two mechanisms, are discussed and compared with published
experimental auditory
nerve data.
\end{abstract}


\maketitle

\section{Introduction}
\label{sec:introduction}

The natural environment presents the auditory system with the
challenge of responding to sounds over many orders of magnitude; the
threshold of hearing and the threshold of pain differ by about
thirteen orders of magnitude.  For the ear to discriminate between
sounds over such a large dynamic range, it is necessary for auditory
stimuli to be compressed into a much smaller, more achievable range of
physical responses.  This is accomplished through a nonlinear
mechanism in which small amplitude sounds are amplified to a greater
extent than larger amplitude sounds.  In order to process complex
sound stimuli, it is also necessary for the auditory system to
distinguish between the frequency components of the stimuli.  The
ear's amplification and frequency discrimination properties are
thought to be derived from a common, metabolically-powered mechanism,
the details of which have been the topic of much investigation
\cite{RR01,M01,FRH01}.

In both mammals and nonmammals, hair cells of the inner ear are
responsible for translating sound-induced mechanical stimuli into a
neurotransmitter signal which induces the firing of the auditory nerve
\cite{Kandel,H89}.  Each hair cell consists of a cell body which is
contacted from below by the auditory nerve and a hair bundle
consisting of actin-supported fibers.  When sound stimulates the
auditory organ, the resulting motion of the hair bundle causes
transduction channels to be mechanically pulled open.  Ionic current
then enters the cell body through the transduction channels, thereby
depolarizing the cell, and ultimately causing the release of
neurotransmitter at the auditory nerve synapse.

In mammals, the frequency-discrimination properties of the basilar
membrane, the membrane in which the hair cells are embedded,
contribute to the auditory system's capacity to distinguish between
sounds of different frequencies.  By contrast in nonmammals, the
surface in which the hair cells are embedded lacks tuning properties.
The nonmammalian auditory system is thought to achieve its frequency
tuning properties through two different mechanisms, both intrinsic to
the hair cell.  The first mechanism involves the mechanical motion of
the hair bundle. Experiments indicate that the hair bundle responds
actively, with greater energy than provided by the stimulus, if forced
near its resonance frequency \cite{MH99}.  
Evidence for a second mechanism,
referred to as electrical resonance, is provided by the
decaying oscillations that are observed in the membrane
potential of the cell body in response to constant current injection
\cite{FC78}.  These oscillations indicate 
that the cell body possesses a preferred response frequency.

Dynamical systems methods have proven useful in analyzing the
frequency tuning and amplification properties of physiologically-based
auditory models. Interestingly, models of both the active motion of
the hair bundle and the electrical resonance mechanism have been shown
to be poised near a Hopf bifurcation~\cite{CMH98,OEM01}. A Hopf
bifurcation is a robust mechanism for generating spontaneous
oscillations as a control parameter of a nonlinear system is varied. It
occurs when a static equilibrium loses stability via a complex
conjugate pair of eigenvalues (of the associated linear stability
problem) crossing the imaginary axis in the complex plane with nonzero 
imaginary part. It has been suggested that the hair cell critically tunes
itself so that its parameters are poised just below the bifurcation point,
thereby making the cell sensitive to stimuli at the Hopf bifurcation
frequency, without causing spontaneous oscillations \cite{EOCHM00,CDJP00}.  
These investigations determine the generic frequency tuning and
amplification properties of a periodically-forced system in the
vicinity of a Hopf bifurcation; specifically, they analyze 
the characteristics of
solutions that are frequency-locked
to a weak, additive resonant forcing term.  Sufficiently close to the
Hopf bifurcation point, the system is compressively nonlinear: small
inputs are amplified to a greater extent than larger
ones~\cite{EOCHM00,CDJP00}. Moreover, the compression of the dynamic
range is accompanied by frequency tuning, which is sharper for small amplitude
inputs than for larger amplitude signals.

Previous studies of hair cell amplification models have considered the
normal form for a system near a Hopf bifurcation without actually
performing the normal form reduction from the physiologically relevant
mathematical model.  Here, in Appendix II, we reduce the Hudspeth and
Lewis model of the electrical tuning mechanism~\cite{HL88a,HL88b} to
the normal form for a system near a Hopf bifurcation, thereby
determining the numerical values of the coefficients of the normal form
corresponding to the model and parameters used by Hudspeth and Lewis. 
We find that the coefficient
of the nonlinear term in the normal form has comparable real and
imaginary parts; it is not purely real as was assumed in earlier
investigations~\cite{EOCHM00,CDJP00}. We show that a result of the
nonzero imaginary part is that the response of the system to
resonant forcing may be hysteretic and that the frequency tuning
curves are no longer symmetric about the resonance frequency.
We further propose a model that describes weak coupling between the
hair bundle amplification mechanism and the electrical resonance
mechanism.  We assume that both oscillation mechanisms are critically tuned to
approximately the same resonant frequency and only weakly damped so
that the Hopf bifurcation normal form applies to each independently,
{\it i.e.} when the other mechanism is suppressed. We then assume weak
linear coupling of the mechanisms, and direct forcing of the hair
bundle at a frequency that is close to its natural frequency. As in earlier 
investigations, the
analysis focuses on the frequency-locked solutions, and, in
particular, on how the magnitude of response grows with the
forcing. We find that the combined critically-tuned amplification
system, can lead to a response, $R$, that scales with $F^{1/9}$, where
$F$ is the resonant forcing amplitude, thereby leading to enhanced
amplification, $R/F$, of small signals. We also explore the enhanced
frequency-tuning characteristics of the combined mechanical and
electrical amplification system, comparing it with those associated
with a single tuning mechanism.

Our paper is organized as follows. In section II we introduce the
reduced mathematical model, with the mathematical details of the
reduction from the physiologically-detailed models relegated to
Appendix II. Section III contains our analysis of the reduced
model, focusing particularly on the simpler situation of
unidirectional coupling from the hair bundle tuning mechanism to the
electrical resonator. We present response-versus-forcing curves that
demonstrate the transition from a linear response ($R\propto F$) to a
response $R\propto F^{1/9}$ as the amplitude of the (weak) signal
increases. We also demonstrate the sharper tuning that is possible
with the combined amplification system. Finally, section IV compares
our model predictions with published experimental
auditory nerve data.

\section{Model}
\label{model}

Two distinct mechanisms contribute to auditory tuning in nonmammalian
vertebrates: an `electrical' resonance arising from an interplay
between ionic currents through the cell membrane, and a `mechanical'
resonance associated with the active motion of the stereocilia in
reponse to stimuli at their resonance frequency.  We start our
discussion by focusing on the electrical resonance mechanism. The
underlying biophysical components have been discussed by Hudspeth and
Lewis \cite{HL88a,HL88b}, who performed a set of experiments that
carefully characterized the dynamical properties of the major ion
channels on the cell bodies of bullfrog saccular hair cells.  On the
basis of these experiments, they developed a single compartment model
of the hair cell (see Appendix I) using the simplifying assumption
that only two major active ion channels, a voltage-gated calcium
channel and a calcium-gated potassium channel, contribute to the
cell's dynamical behavior.  In this model, the dynamical evolution of
the membrane potential $V_m$ is given by an equation based on the
direct application of Kirchoff's laws to a circuit that represents the
flow of ions across the membrane:

\begin{equation}
- C_{m} \frac{dV_{m}}{dt} = g_{Ca} m^3 (V_m-E_{Ca})
                          +g_{K(Ca)}(O_2+O_3)(V_m - E_K) 
                            + g_L (V_m - E_L) - I.
\label{eq:Veq1}
\end{equation}
Here $V_{m}$ is the membrane potential and $C_{m}$ is the membrane
capacitance per unit area.  The voltage-gated calcium ($\textrm{Ca}$)
current is represented by $g_{\textrm{ Ca}} m^3 (V_m-E_{\textrm{Ca}})$,
where $g_{\textrm{Ca}}$ is the maximum $\textrm{Ca}$ conductance per
unit area, $m$ is the voltage-dependent fraction of open
conformational subunits in the $\textrm{Ca}$ channels, and
$E_{\textrm{Ca}}$ is the reversal potential for the $\textrm{Ca}$ ion
channels. The $\textrm{Ca}$-gated potassium ($\textrm{K}$) current is
represented by $g_{K(\textrm{Ca})}(O_2+O_3)(V_m - E_\textrm{K})$,
where $g_{\textrm{K(Ca)}}$ is the maximum $\textrm{K}$ conductance per
unit area, $(O_2+O_3)$ is the fraction of $\textrm{K}$ channels in one
of their two open states, and $E_\textrm{K}$ is the reversal potential
for the $\textrm{K}$ ion channels. The term $g_L (V_m - E_L)$
represents all passive ion channels as a leak conductance $g_L$ per
unit area and a reversal potential $E_L$. The command current $I$ is
directly injected into the cell body. The formulation of the model
involves six additional equations (see Appendix I) that describe the
dynamical evolution of the fraction $m$ of open units in the
$\textrm{Ca}$ channels, the intracellular concentration of
$\textrm{Ca}$ ions close to the cell membrane, and the fraction of
$\textrm{Ca} $-gated $\textrm{K}$ channels in each of their three
closed states $(C_0, C_1, C_2)$ and two open states $(O_2, O_3)$.  In
their seminal work \cite{HL88a,HL88b}, Hudspeth and Lewis (H\&L)
experimentally characterized the value of the various parameters that
appear in these equations.

\begin{figure}
\centerline{\resizebox{3 in}{!}{\includegraphics{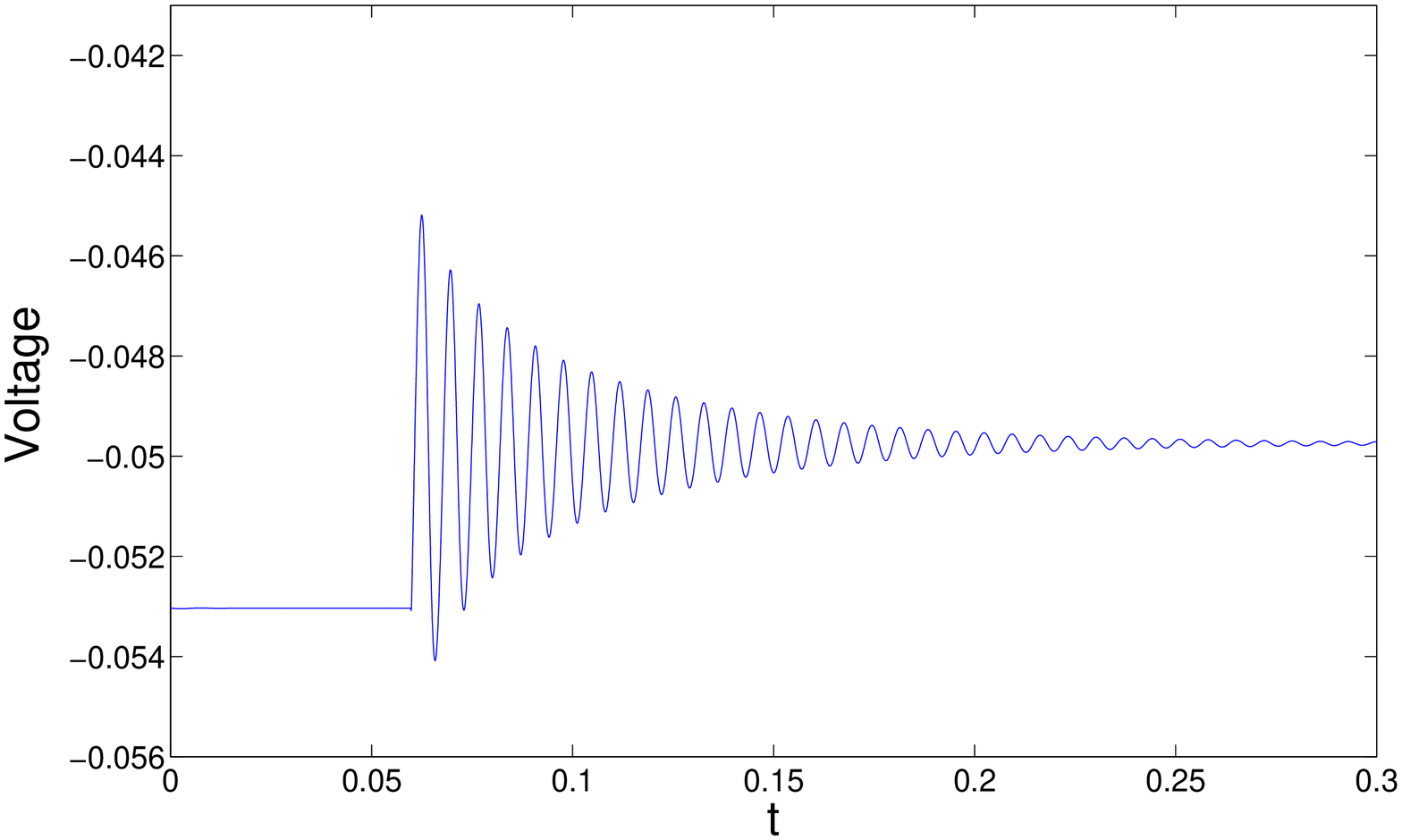}}
\resizebox{3 in}{!}{\includegraphics{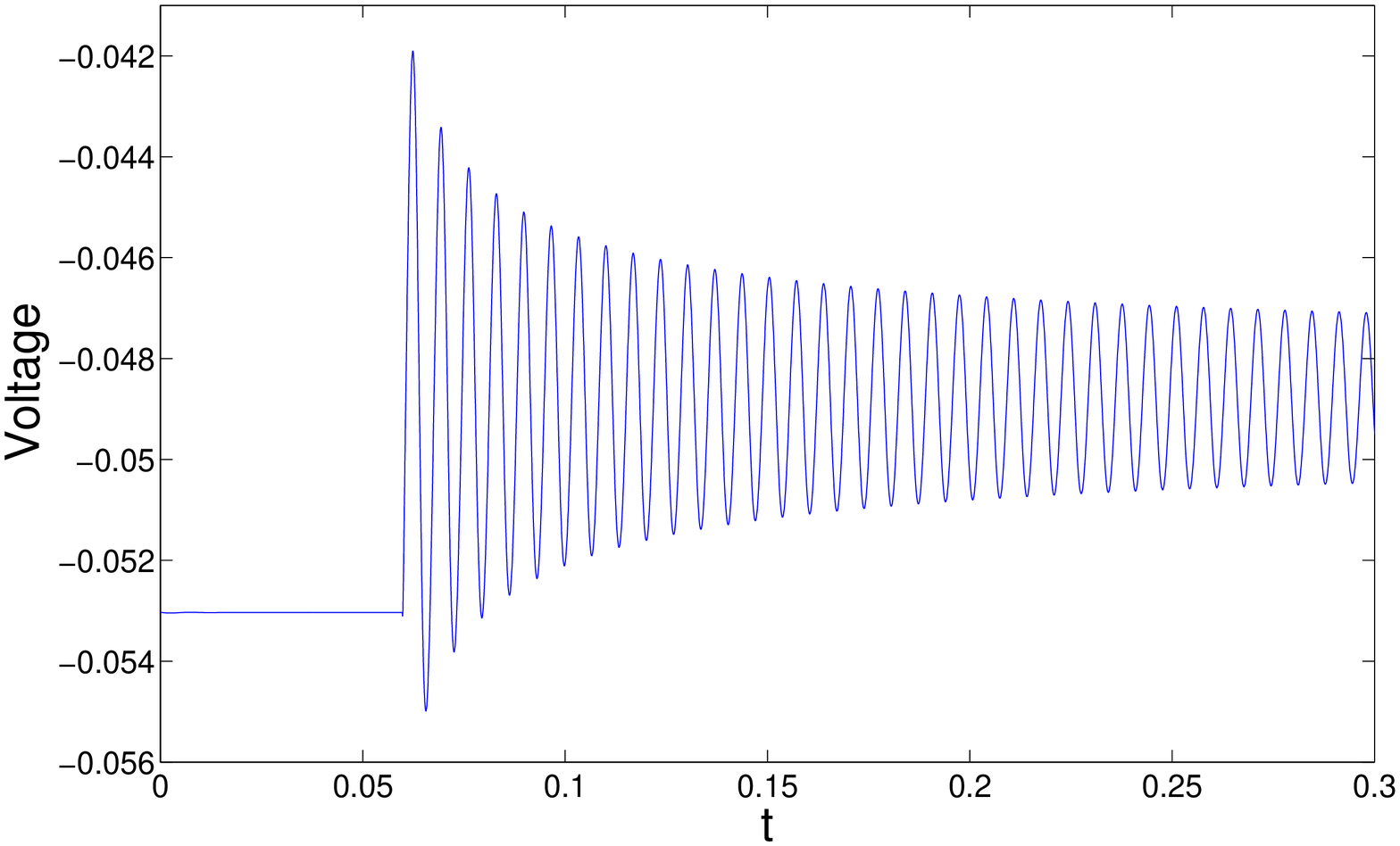}}}
\centerline{\hspace{.1 in} (a) \hspace{2.7 in} (b)}
\caption{Numerical simulations of the response of the membrane
potential of a hair cell in the Hudspeth and Lewis
model~\cite{HL88a,HL88b}, to a constant current, $I$, injected at
$t=.06$.  (a) $I=65$ pA; (b) $I=95$ pA. A Hopf bifurcation occurs at
$I^{*} \approx 91.3$ pA. The other parameters of the H\&L
model~(\ref{eq:hlmodel}), used in the simulations, are given in
Appendix I.}
\label{fig:hudplots}
\end{figure}

The H\&L model reproduces qualitatively the decaying membrane
potential oscillations observed in current-clamp experiments in which
a current of constant amplitude $I$ is injected into the cell body
\cite{HL88a,HL88b}. As shown in Figure~\ref{fig:hudplots}, the
response of the membrane potential to a step current of amplitude $I$
depends crucially on the value of the injected current. For small
injected currents, as illustrated in Figure~\ref{fig:hudplots}(a), the
membrane potential exhibits an oscillatory decay to a new constant
value. For larger current values, as illustrated in
Figure~\ref{fig:hudplots}(b), the membrane potential decays to a new
state that is oscillatory. This qualitative difference signals a
transition between a regime in which the asymptotic state is a fixed
point and a regime in which the asymptotic state is a limit cycle. In
the H\&L model, this transition occurs by a Hopf bifurcation
\cite{OEM01}.

\begin{figure}
\centerline{\resizebox{3.6 in}{!}{\includegraphics{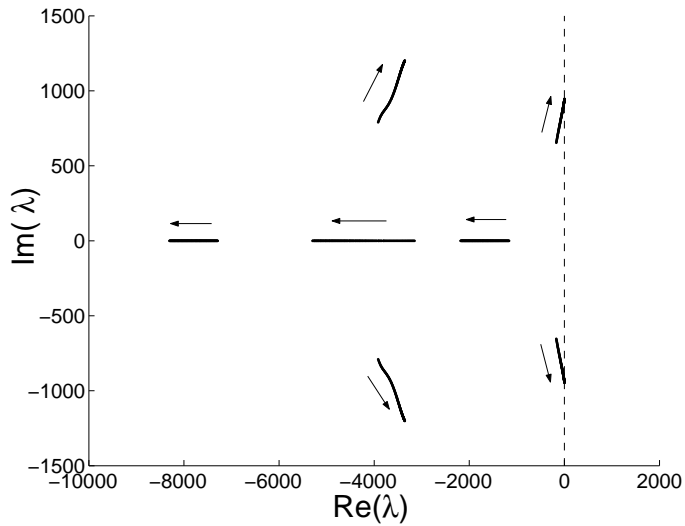}}}
\caption{The eigenvalues of the Hudspeth and Lewis
model~(\ref{eq:hlmodel}), linearized about its static equilibrium
state as described in Appendix II, evolve in the complex plane as $I$ is
increased from 0 pA to 100 pA.  Solid lines indicate the trajectory of
the eigenvalues with increasing $I$. Other parameters of the equations
are given in Appendix I.}
\label{fig:hudeigs}
\end{figure}

A Hopf bifurcation occurs when a fixed point of a system of ODE's
undergoes a change in stability in which a complex conjugate pair of
eigenvalues $\lambda$, $\overline{\lambda}$ passes from the
$Re(\lambda)<0$ to the $Re(\lambda)>0$ side of the imaginary axis in
the complex plane.  Figure~\ref{fig:hudeigs} shows the evolution of
the eigenvalues of the H\&L model linearized around its fixed point,
as the input current is increased from 0 and 100 pA. Note that three
of the eigenvalues are always real and negative, and one complex
conjugate pair remains in the $Re(\lambda)<0$ semispace. The leading
complex conjugate pair crosses the $Re(\lambda)=0$ axis for $I^{*}
\approx 91.3$ pA; this is the value of the input current at which the
fixed point becomes unstable, as determined by fixing the parameters
of the model at the experimentally--based estimates listed in Hudspeth
and Lewis's original paper.  That the H\&L model is, for
physiologically reasonable values of the parameters, poised near a
Hopf bifurcation has profound implications for the signal processing
capabilities of the modelled hair cell. Specifically, at the Hopf
bifurcation the system is compressively nonlinear, as it exhibits a
large amplification of small amplitude inputs and a smaller amplification of
large amplitude inputs \cite{EOCHM00,CDJP00}. Moreover, the resulting
compression of the dynamic range is accompanied by a sharp frequency
tuning of the response to small amplitude inputs and a broad tuning in the
response to large amplitude inputs \cite{EOCHM00,CDJP00}.

The dynamical behavior of the model is asymptotically described by the
amplitude of the mode associated with the most unstable eigenvector,
which is the one associated with the complex conjugate pair of
eigenvalues that cross the $Re(\lambda)=0$ axis at the Hopf
bifurcation.  Sufficiently close to the  bifurcation, the system
can be generically described by a normal form equation~\cite{W1990}:

\begin{equation}
\frac{dA}{dt}=(a +  ib) A +(c +  id) |A|^2 A, 
\label{eq:HopfNormal}
\end{equation}

\noindent 
which characterizes the dynamical evolution of the complex amplitude
$A$ of the most unstable mode. In this equation, the parameter $a$ is
a measure of the distance to the Hopf bifurcation at $a=0$; $a$ is
negative below the bifurcation and positive above.  The parameter $b$
is the frequency of the system at the bifurcation, and the parameter
$d$ measures the shift in preferred frequency as the amplitude of the
solution increases, as readily seen by setting $A=re^{i\Omega t}$
in~(\ref{eq:HopfNormal}) to establish that $r=\sqrt{{a\over -c}}$ and
$\Omega=b+dr^2$.  The parameter $c$ distinguishes between
supercritical $(c<0)$ and subcritical $(c>0)$ Hopf bifurcations.

We have carried out a standard nonlinear reduction of the H\&L model
to the normal form (see Appendix II). In the resulting normal form
(\ref{eq:finalresult}), $a$ is proportional to $\Delta I = (I -
I^{*})/ I^{*}$ and $c$ is negative. This reduction thus establishes
the supercritical nature of the Hopf bifurcation in the H\&L model.
For this supercritical bifurcation, there is a transition from a
stable fixed point to a state in which the fixed point becomes
unstable and a stable limit cycle exists. Moreover, in the
supercritical case, the radius $r$ of the limit cycle grows, with
distance past the bifurcation point, with the characteristic scaling
$r\propto \sqrt{\Delta I}$. 

In the current-clamp experiments conducted by Hudspeth and Lewis
\cite{HL88a,HL88b}, the current injected into the cell was of constant
amplitude. In contrast, natural inputs to the cell are due to a
time-dependent, sound-induced mechanical displacement of the hair
bundle, which results in time-dependent changes in the conductance
through the transduction channels.
Because the transduction channels are passive, this 
time-dependence can be incorporated into the H\&L model through
the leakage conductance term on the right-hand-side
of~(\ref{eq:Veq1}). In the case of a simple time-periodic conductance,
the reduction carried out in Appendix II suggests that for command
currents close to $I^*$ the most important contribution of the
time-periodic forcing to the asymptotic dynamics comes from the
Fourier component that is closest to the resonance frequency of the
system. In the case of a weak periodic signal with frequency close to
the resonator's frequency, the time-periodic input can be represented
as an additive contribution to the amplitude equation, which then
takes the form:
\begin{equation}
   \frac{dA}{dt} = (a + i b ) A +(c+id)|A|^2 A + F e^{i \omega t}.
\label{eq:singleredf}
\end{equation}
Due to the time-translation symmetry of the unforced case, we can,
without loss of generality, assume that $F$ is real and positive.
We note that earlier investigations of~(\ref{eq:singleredf}) in the context of
amplification mechanisms in auditory hair cells~\cite{EOCHM00,CDJP00}
assumed a real coefficient of the nonlinear term, {\it i.e.} they made
the nongeneric assumption that $d=0$ such that the preferred frequency of the nonlinear
oscillator had no amplitude-dependence.

We next consider the tuning mechanism associated with the mechanical
deflection of the hair bundle.  Experiments in which a glass fiber was
attached to the hair bundle and used to mechanically stimulate the
bundle at a specific frequency showed that the hair bundles respond
preferentially to stimuli at their resonance frequency \cite{MH99}.  
The motion of the hair bundles has been shown to be sensitive to
the amount of calcium ion entering the transduction channels \cite{CC06}. 
A model for hair bundle motion due to calcium binding 
within the stereocilia was proposed by Choe {\it et~al}.~\cite{CMH98}. 
In their model, when transduction current enters
the hair bundle $ \textrm{Ca}$ ions attach themselves to the
transduction channels at sites within the stereocilia; this attachment
causes an increase in tension, which in turn causes the channel to
close \cite{CMH98}. When the transduction channel closes, the
$\textrm{Ca}$ ions detach themselves from the binding sites and the
channel returns to its regular tension allowing the cycle to repeat
itself.  Choe {\it et al}. have shown that, when the parameters of their
model are specified within physiologically reasonable ranges, the
model is near a Hopf bifurcation.  Experimental evidence 
also indicates that the relationship between stimulus magnitude 
and magnitude of the hair bundle oscillations obeys
the scaling that would be expected for a system tuned near a Hopf
bifurcation \cite{MH01}, lending support to our assumption that a
model describing the hair bundle dynamics is poised near a Hopf
bifurcation.  Assuming this second mechanism is tuned sufficiently
close to a Hopf bifurcation, it too can be reduced to the normal
form~(\ref{eq:HopfNormal}) for a system near a Hopf bifurcation.

It only remains to consider first the manner in which the two tuning
mechanisms are coupled in the biological system, and second how this
coupling should be represented in the reduced model. One clear source
of coupling between the two tuning mechanisms is through the
transduction current.  The magnitude of the transduction current
entering through the stereocilia is directly related to the magnitude
of displacement of the stereocilia.  This relationship between
displacement and the amount of current entering the cell has been
measured indirectly by measuring the change in the receptor potential
of the cell in response to stereocilia displacements of different
magnitudes \cite{HC77}.  Such experiments indicate that, in absence of
stimuli, a small amount of current is flowing into the cell through
the stereocilia.  When the hair bundle is deflected in the negative
direction, away from the tallest stereocilia, transduction channels
close and the amount of current flowing into the cell decreases.
Similarly, when the hair bundle is deflected in the positive
direction, the amount of current entering through the transduction
channels increases and eventually saturates. For small displacements
in the positive direction, the relationship between the displacement
of the hair bundle and the change in the receptor potential is
approximately linear \cite{HC77}.  As the amplitude of the hair bundle
oscillations increases due to stimulation at the stereocilia's
resonance frequency, the amount of current entering the cell body and
providing a forcing to the second electrical resonance mechanism
increases.  This clearly provides a means of coupling from the
stereocilia tuning mechanism to the electrical resonance mechanism.

There is also evidence for coupling in the reverse direction, from
the electrical resonance mechanism to the hair bundle resonance
mechanism, in that electrical stimulation of the cell body has been
shown to induce displacement of the stereocilia
\cite{AC92,DW92,RCF00,RCF02,BH03}.  The exact mechanism for coupling
in this direction is less clear. Experimental comparisons between
current injected into the cell body and the resulting displacement of
the stereocilia indicate that the linear approximation is reasonable
for small current injections \cite{RT90}.  The presence of coupling in
both directions raises the question of whether there are actually two
separate tuning mechanisms or the stereocilia tuning mechanism is
simply a manifestation of the electrical tuning mechanism.  This
question was addressed in experiments in which the cell body was
voltage clamped to silence electrical resonance oscillations allowing
the motion of the stereocilia to be probed separately \cite{MBCH03}.
These experiments demonstrate active motion of the stereocilia even in
the absence of the electrical resonance mechanism.  Electrical resonance
experiments are often performed by direct current injection, with
transduction channels blocked, so the independence of the electrical
resonance mechanism from the active motion of the stereocilia was
never in question.

From the biological evidence, it is reasonable to assume that the
coupling between the two mechanisms is linear for sufficiently small
forcing.  This leads
us to a reduced model consisting of two coupled amplitude equations of
the form,

\begin{equation}
  \frac{dA_1}{dt}=(a_1 + i b_1 ) A_1 +(c_1 + i d_1 ) |A_1|^2 A_1 + \gamma_1 e^{i\psi_1} 
A_2 + F e^{i \omega t},
\label{eq:coupledampa}
\end{equation}

\begin{equation}
  \frac{dA_2}{dt}=(a_2 + i b_2 ) A_2 +(c_2 + i d_2 ) |A_2|^2 A_2 + \gamma_2 e^{i\psi_2}A_1.
\label{eq:coupledampb}
\end{equation}
In this model, (\ref{eq:coupledampa}) represents the hair bundle
resonance mechanism which receives a sound-induced time-dependent
forcing ($\propto F$) as well as feedback from the electrical
resonance mechanism ($\propto A_2$). Equation~(\ref{eq:coupledampb})
represents the electrical resonance mechanism which receives a forcing
proportional to the displacement of the stereocilia.  Note that we
allowed for a phase difference $\psi_j$ in each of the coupling terms,
with the corresponding parameters $\gamma_j$ taken to be real and
non-negative. Moreover, we note that when both $\gamma_1$ and
$\gamma_2$ are nonzero, then the Hopf bifurcations that cause
spontaneous oscillations in the unforced problem ($F=0$) will shift
away from $a_j=0$. As detailed mathematically in Appendix II, the
model is valid for the case in which each system is tuned close to the
Hopf bifurcation ($|a_j|$ sufficiently small), each system is tuned
near the resonance frequency ($b_j \approx \omega$), and the forcing
and the coupling are weak ($F$, $\gamma_j$ sufficiently small).

We have determined from Hudspeth and Lewis's model the numerical
values of the coefficients $a_2,\ b_2,\ c_2,\ d_2$ of
(\ref{eq:coupledampb}) for H\&L's physiological parameters.  Models
also exist for the stereocilia mechanism, so it is possible that the
same coefficients of~(\ref{eq:coupledampa}) could be determined based on
these models. Performing the second reduction however would not be
particularly useful in this paper because our analysis requires that
both systems be tuned close to the same frequency.  There are multiple
ways to tune the physiological models such that they yield vibrations
at a required frequency.  Thus, without a very clear idea of the
physiologically reasonable method to adjust the model parameters,
there is no way to determine a consistent result for the numerical
values of the coefficients in the amplitude
equation~(\ref{eq:coupledampa}).  Nonetheless, many of our conclusions
regarding scaling laws hold provided that the Hopf bifurcations are
supercritical ({\it i.e.} provided $c_j<0$), and that our fundamental
modeling assumptions are met.

\section{Analysis}
\label{analysis}

\subsection{Response-Versus-Forcing Relationship}
\label{sec:unidirectional}

Our analysis of the reduced model,
Eqs.~(\ref{eq:coupledampa})-(\ref{eq:coupledampb}), focuses on
frequency-locked solutions of the form $A_j = R_j e^{i(\omega t +
\phi_j)}$, $j=1,2$, where $R_j \geq 0$ and $\phi_j\in [0,2 \pi)$ are
constants. We wish to determine how the magnitude of the electrical
response, measured by $R_2$, scales with the sound-induced mechanical
forcing amplitude $F$.  This scaling depends on the proximity to the
Hopf bifurcation, captured by the linear damping coefficients $a_j<0$
in (\ref{eq:coupledampa})-(\ref{eq:coupledampb}). It also depends on
how closely tuned the natural frequencies, $b_j$, of the nonlinear
oscillation mechanisms are to each other and to the driving frequency,
$\omega$.

Substituting $A_j=R_je^{i(\omega t+\phi_j)}$
into (\ref{eq:coupledampa})-(\ref{eq:coupledampb}) yields the following
pair of complex-valued algebraic equations defining an implicit relationship
between the real quantities $F$ and $R_2$:
\begin{eqnarray}
  F e^{-i\phi_1} &=&-(a_1+i(b_1-\omega))R_1-(c_1+id_1)R_1^3
 -\gamma_1 R_2e^{i(\phi_2-\phi_1+\psi_1)},\label{eq:m1} \\
  \gamma_2 R_1 e^{i(\phi_1-\phi_2+\psi_2)}&=&-(a_2+i(b_2-\omega))R_2
-(c_2+id_2)R_2^3.
\label{eq:m3} 
\end{eqnarray}
We can solve~(\ref{eq:m3}) for $R_1$ in terms of $R_2$:
\begin{equation}
  R_1 = -\frac{e^{-i(\phi_1-\phi_2+\psi_2)}}{\gamma_2}
\Bigl((a_2+i(b_2-\omega))R_2
+(c_2+id_2)R_2^3\Bigr)\ .
\label{eq:R1}
\end{equation}
Here we can determine the
phase difference $(\phi_1-\phi_2)$ by
the requirement that $R_1$ be real and nonnegative.
Substituting this expression into~(\ref{eq:m1}), we find
\begin{equation}
F=e^{i(\phi_2-\psi_2)}
\Bigl(\alpha_1 R_2+\alpha_3 R_2^3+\alpha_5 R_2^5+\alpha_7 R_2^7
+\alpha_9 R_2^9
\Bigr),
\label{eq:poly}
\end{equation}
where 
\begin{eqnarray}
\alpha_1&\equiv&
-\gamma_1e^{i(\psi_1+\psi_2)}
+\frac{1}{\gamma_2}(a_1+i(b_1-\omega))(a_2+i(b_2-\omega)),
\nonumber\\
\alpha_3&\equiv&\frac{1}{\gamma_2}(c_2+id_2)(a_1+i(b_1-\omega))
+\frac{1}{\gamma_2^3}e^{-2i(\phi_1-\phi_2+\psi_2)}
(c_1+id_1)(a_2+i(b_2-\omega))^3,\nonumber\\
\alpha_5&\equiv&\frac{3}{\gamma_2^3}
e^{-2i(\phi_1-\phi_2+\psi_2)}(a_2+i(b_2-\omega))^2
(c_1+id_1)(c_2+id_2)
,\label{eq:alphacoeffs}\\
\alpha_7&\equiv&
\frac{3}{\gamma_2^3}e^{-2i(\phi_1-\phi_2+\psi_2)}(a_2+i(b_2-\omega))(c_1+id_1)
(c_2+id_2)^2
,\nonumber\\
\alpha_9&\equiv&\frac{1}{\gamma_2^3}e^{-2i(\phi_1-\phi_2+\psi_2)}(c_1+id_1)
(c_2+id_2)^3\ .\nonumber
\end{eqnarray}
Again, the phase $\phi_2$ in~(\ref{eq:poly}) is determined by the
requirement that the forcing magnitude $F$ be real and non-negative.
It immediately follows from the polynomial form of~(\ref{eq:poly})
that the response $R_2$ need not be a single-valued function of $F$.
It also follows that, with increasing forcing, there is a
transition from a linear regime, $R_2\propto F$, for sufficiently
small forcing, to a regime where $R_2\propto F^{1/9}$ for larger values
of $F$.  However, whether this transition occurs for small values of
$F$, for which the model~(\ref{eq:coupledampa})-(\ref{eq:coupledampb}) 
is valid, depends on
both the magnitude of the coupling coefficients $\gamma_j$ and the
magnitude of the linear coefficients $a_j+i(b_j-\omega)$. In
particular, if we let $\epsilon=|a_2+i(b_2-\omega)|$, then we expect the transition to the regime $R_2\propto
F^{1/9}$ will occur for small $F$ provided that
$|a_1+i(b_1-\omega)|\gamma_2^2$ is at most {\cal O}$(\epsilon^3)$, and
$\gamma_1\gamma_2^3$ is at most {\cal O}$(\epsilon^4)$.

\begin{figure}
\centerline{\resizebox{3in}{!}{\includegraphics{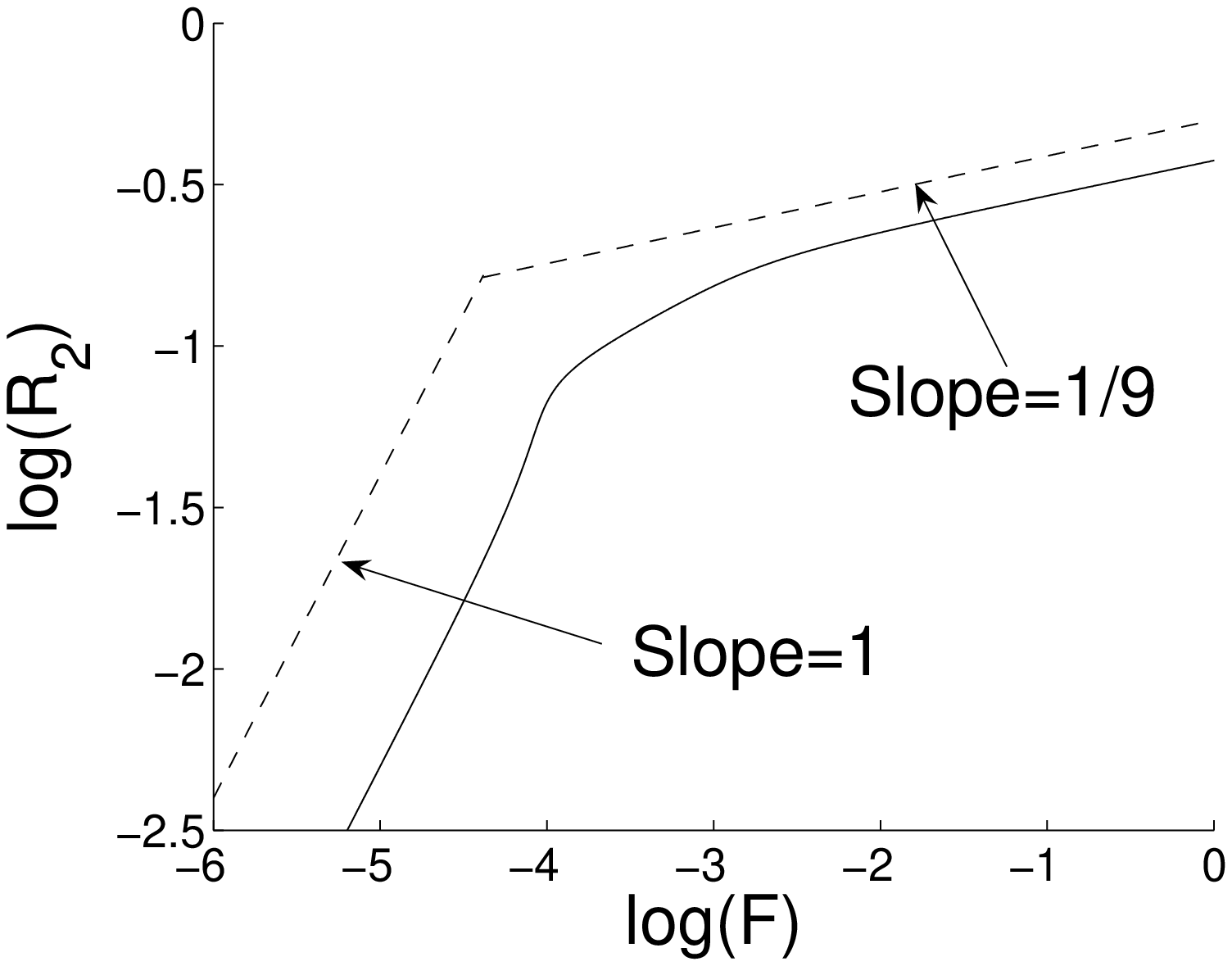}}
\resizebox{3 in}{!}{\includegraphics{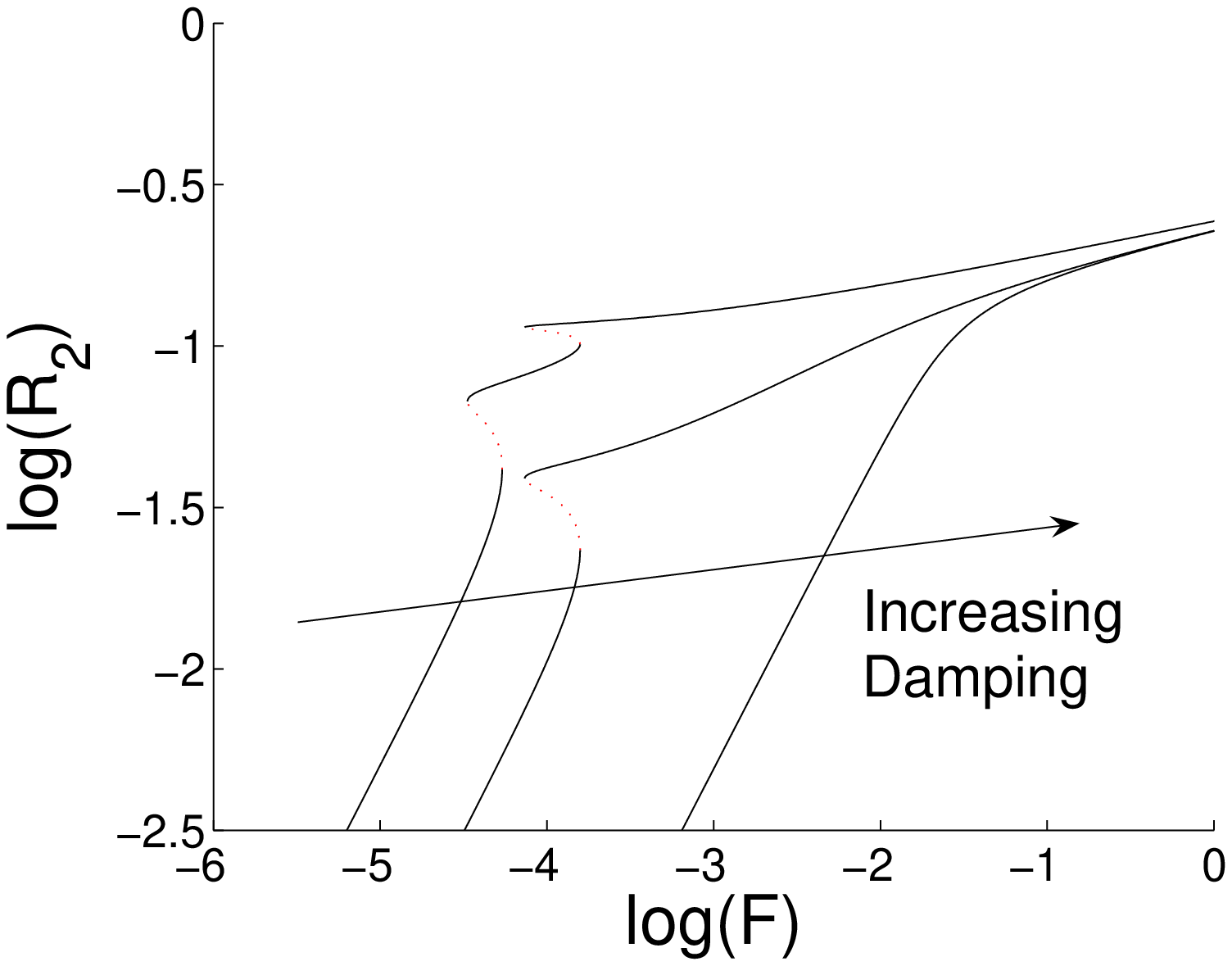}}}
\hspace{.1 in} (a) \hspace{2.8 in} (b)
\centerline{\resizebox{3 in}{!}{\includegraphics{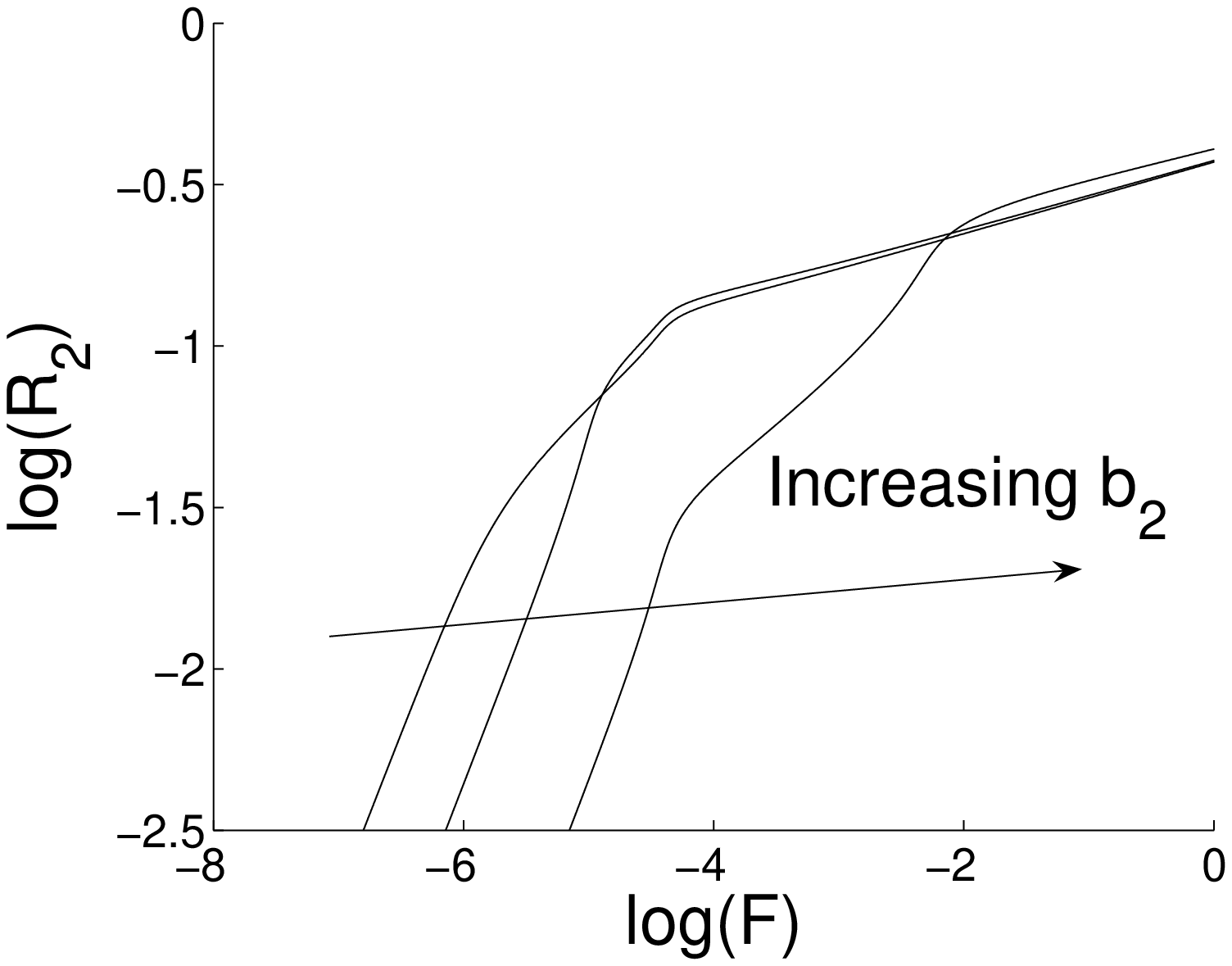}}
\resizebox{3 in}{!}{\includegraphics{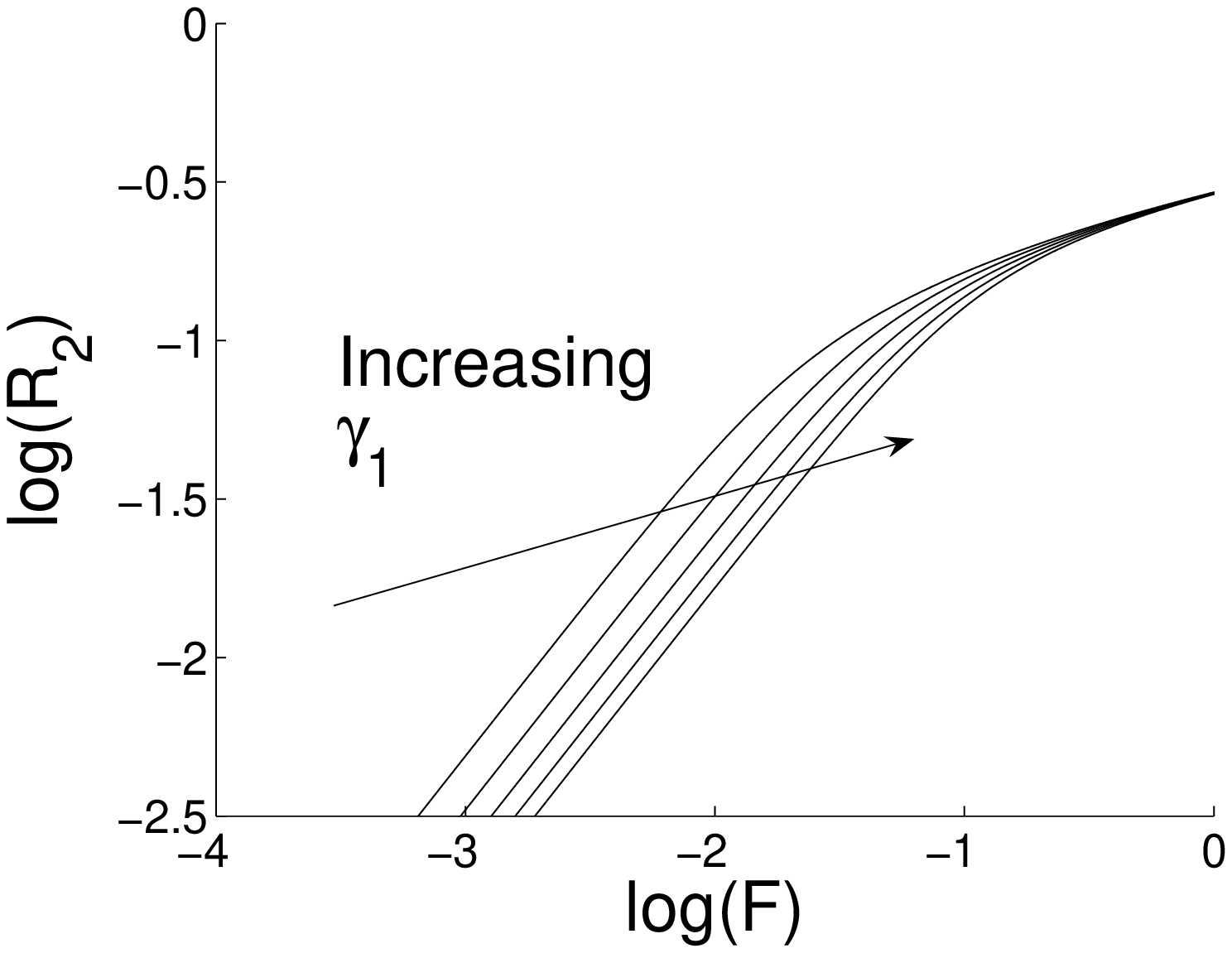}}}
\hspace{.1 in} (c) \hspace{2.8 in} (d)
\caption[Log-log plot of response ($R_2$) versus forcing ($F$)]{\small
Sample log-log plots of response ($R_2$) versus forcing ($F$) from
(\ref{eq:poly}-\ref{eq:alphacoeffs}), for the system with $c_1=c_2=-1$
and with varying damping, coupling, and detuning magnitudes.  Solid
(dotted) lines represent (un)stable frequency-locked solutions.  (a)
Transition from linear response ($R_2\propto F$) to nonlinear response
with $R\propto F^{1/9}$; dashed lines, for comparison, have slopes 1
and 1/9 as indicated.  Parameters set to
$a_1=-0.02$, $a_2=-0.001$, $b_1-\omega=0$,
$b_2-\omega=.01$, $d_1=-2$, $d_2=-1.1$,
$\gamma_1=0$, $\gamma_2=.1$, and $\psi_2=0.64\pi$.  
(b) Response curves obtained with linear damping parameters
$(a_1,a_2)=(-.0002,-.0001),(-.0002,-.1),(-.2,-.1)$;  other
parameters set at $b_1-\omega=.01$, $b_2-\omega=.02$, 
$d_1=-6$, $d_2=-4$, $\gamma_1=0$, $\gamma_2=.1$, 
and $\psi_2=0.64\pi$.  (c) Response curves obtained with
detuning $b_2-\omega=.001,.01,.1$;  other parameters set at
$b_1-\omega=.002$, $a_1=-.001$, $a_2=-.002$,
$d_1=-2$, $d_2=-1.1$, $\gamma_1=0$, $\gamma_2=.1$, and
$\psi_2=0.64\pi$.  (d) Response curves obtained with different backward
coupling magnitudes of $\gamma_1=0, .1, .2, .3, .4$; other parameters set
at $a_1=-0.1$ $a_2=-0.2$, $b_1-\omega=0.02$, $b_2-\omega=0.01$,
$d_1=-1$, $d_2=-1.1$, $\gamma_2=.1$, $\psi_1= \pi$, $\psi_2=0$.}
\label{fig:RFplots}
\end{figure}

The response {\it vs.} forcing characteristics associated with the
model~(\ref{eq:coupledampa})-(\ref{eq:coupledampb}) are further
explored in Fig.  \ref{fig:RFplots}.  For this figure, we assume
supercritical Hopf bifurcations associated with the mechanical and
electrical resonance mechanisms so that $c_j<0$ in the model
equations. With an appropriate re-scaling of amplitudes $A_j$, we may
then assume $c_j=-1$ for both $j=1$ and $j=2$.  Our direct calculation
of the normal form coefficients in~(\ref{eq:coupledampb}) (see
Appendix II), from the H\&L model, yields $d_2/c_2\approx 1.1$. Thus
in many of our numerical computations we set $d_2=-1.1$. Finally, if
we scale time by the dimensioned forcing frequency, we may take
$\omega=1$ in the model equations. 
Fig.~\ref{fig:RFplots} indicates stable frequency-locked solutions as
solid lines and unstable solutions by dotted lines. The linear
stability of the frequency-locked solutions is determined by
substituting the ansatz $A_j=R_je^{i(\omega t+\phi_j)}(1+z_j(t))$ into
Eqs.~(\ref{eq:coupledampa})-(\ref{eq:coupledampb}) and then
linearizing about $z_j=0$. We then find that the perturbations $z_j$
satisfy the following system of linear differential equations

\begin{equation}
\left(\begin{array}{c} \dot{z}_1\\ \dot{\overline{z}}_1 
\\ \dot{z}_2\\ \dot{\overline{z}}_2\end{array}\right)
=\left(\begin{array}{cccc}
M_1 & M_2 & M_3 & 0\\
\overline{M}_2 & \overline{M}_1&0&\overline{M}_3\\
M_6 & 0 & M_4 & M_5\\
0 & \overline{M}_6 &\overline{M}_5 &\overline{M}_4
\end{array}\right)
\left(\begin{array}{c} {z}_1\\ \overline{z}_1 
\\ {z}_2\\ \overline{z}_2\end{array}\right),
\label{eq:linearstab}
\end{equation}
where $\overline{z}_1$ denotes the complex conjugate of $z_1$, {\it etc}.,
and
\begin{eqnarray}
M_1&\equiv& a_1+ib_1-i\omega+2(c_1+id_1)R_1^2 \nonumber\\
M_2&\equiv& (c_1+id_1)R_1^2\nonumber\\
M_3&\equiv& \gamma_1 {R_2\over R_1} e^{i(\phi_2-\phi_1+\psi_1)}\label{eq:matrixelements}\\
M_4&\equiv& a_2+ib_2-i\omega+2(c_2+id_2)R_2^2 \nonumber\\
M_5&\equiv& (c_2+id_2)R_2^2\nonumber\\
M_6&\equiv& \gamma_2 {R_1\over R_2} e^{i(\phi_1-\phi_2+\psi_2)}.\nonumber
\end{eqnarray}
The solution $A_j=R_je^{i(\omega t+i\phi_j)}$, with $R_j$ and $\phi_j$ satisfying
(\ref{eq:m1})-(\ref{eq:m3}), is stable if the eigenvalues of the matrix associated
with the linearized problem (\ref{eq:linearstab}) all have negative
real part.

Figure~\ref{fig:RFplots}(a) demonstrates the predicted transition from
linear response to nonlinear response with $R_2\propto F^{1/9}$. 
Figure~\ref{fig:RFplots}(b) shows that as the damping in the
system increases, larger forcings are necessary to reach this
nonlinear regime. Moreover, this plot demonstrates the possibility of
hysteresis which can occur when $b_j-\omega$ and $d_j$ have opposite
signs, provided the damping is not too
great. Figure~\ref{fig:RFplots}(c)  shows how the transition from
linear to $1/9$ scaling moves to larger forcing when the detuning
$b_2-\omega$ is increased; note that in this plot a scaling of
$R_2\propto F^{1/3}$ is also evident over an intermediate range of
forcings. Changes in the magnitude of the feedback coefficient
$\gamma_1$ can either enhance or degrade the response, $R_2$,
depending upon the coupling phases $\psi_1$ and $\psi_2$.
Figure~\ref{fig:RFplots}(d) shows an example of the change in the
response versus forcing relationship as the feedback coefficient, $\gamma_1$,
is increased.

\subsection{Uniqueness and Stability: Unidirectional Coupling}
\label{sec:stabilitysd}

The problem of determining the uniqueness and stability of solutions
is greatly simplified in the case of unidirectional coupling between
the mechanical and electrical resonators.  In this section we focus
our further analysis on the case where the feedback from the
electrical resonator to the stereocilia can be neglected, {\it i.e.},
we focus on the case $\gamma_1=0$ in~(\ref{eq:coupledampa}).

In this unidirectional-coupling case, $M_3=0$ in the stability
matrix~(\ref{eq:linearstab}), and the linear stability problem
simplifies to one of determining the eigenvalues associated with the
$2\times 2$-blocks on the diagonal. For instance, if we let $\sigma_1$
and $\sigma_2$ be the eigenvalues associated with the ($\dot{z}_1$,
$\dot{\overline z}_1$)-equations, we find
\begin{eqnarray}
\sigma_1+\sigma_2&=&M_1+\overline{M}_1=2(a_1+2c_1 R_1^2)\label{eq:tracedet}\\
\sigma_1\sigma_2&=& |M_1|^2-|M_2|^2=a_1^2+(b_1-\omega)^2+4[a_1c_1+(b_1-\omega)d_1]R_1^2
+3(c_1^2+d_1^2)R_1^4.\nonumber
\end{eqnarray}
Similar equations for the remaining two eigenvalues, associated with
the ($\dot{z}_2$, $\dot{\overline z}_2$)-equations, hold: they are
obtained from~(\ref{eq:tracedet}) by interchanging the 1 and 2
subscripts on its right-hand-side. In the case of supercritical Hopf
bifurcations ($c_j<0$), and damping of spontaneous oscillations
($a_j<0$), the frequency-locked solutions are stable provided
$|M_1|^2-|M_2|^2>0$ and $|M_4|^2-|M_5|^2>0$.  While these conditions
hold for sufficiently small and sufficiently large amplitudes $R_1$
and $R_2$, they may be violated in the intermediate regime if
$a_jc_j+(b_j-\omega)d_j<0$ for either $j=1$ or $j=2$. Since we are
interested in the case that $a_j$ and $c_j$ are both negative, a
necessary condition for a frequency-locked solution to lose stability,
with increase in forcing $F$, is that $b_j-\omega$ and $d_j$ must have
opposite signs.  To gain some insight into this criterion, it is
useful to note that the preferred frequencies, $b_1+d_1 R_1^2$ and
$b_2+d_2 R_2^2$, of each tuning mechanism, in absence of forcing, are
dependent upon response magnitude.  If the natural frequency of the
cell shifts away from the forcing frequency with increasing response
($[b_j-\omega]d_j >0$), then $a_j c_j +(b_j-\omega)d_j > 0$ and the
entrained solution is always stable and unique. 
Instabilities, and their associated hysteresis in the reponse {\it
vs.} forcing curves, can only occur in the unidirectionally-coupled
case if the preferred frequency of at least one of the tuning
mechanisms shifts towards the forcing frequency with increasing
response amplitude ($(b_j-\omega)d_j<0$).  In this case, the system
may jump from a small amplitude response to a larger amplitude one
with an increase in the forcing. This follows from the observation that
the instabilities, if they occur, come in pairs and correspond to
saddle-node bifurcations along the solution branches $R_j(F)$; see
Fig. \ref{fig:RFplots}(b) for examples. 
The necessary {\it and
sufficient} condition for a pair of saddle-node bifurcations to occur
is
\begin{equation}
 a_j c_j +(b_j-\omega) d_j < - \sqrt{\frac{3}{4}(a_j^2 +(b_j-\omega)^2)(c_j^2 +d_j^2)}\ ,  \label{eq:des3}
\end{equation}
for  $j=1$ and/or $j=2$.
Note that the prediction that
hysteresis can occur for larger detunings is a direct result of the
dependence of the cell's preferred frequency on response amplitude,
an effect which was neglected in previous studies for which the
imaginary part of the nonlinear coefficient ($d_{j}$) was
neglected~\cite{EOCHM00,CDJP00}.
 
\section{Results and Discussion}
\label{sec:scaling}

\begin{figure}

\centerline{\resizebox{2 in}{!}{\includegraphics{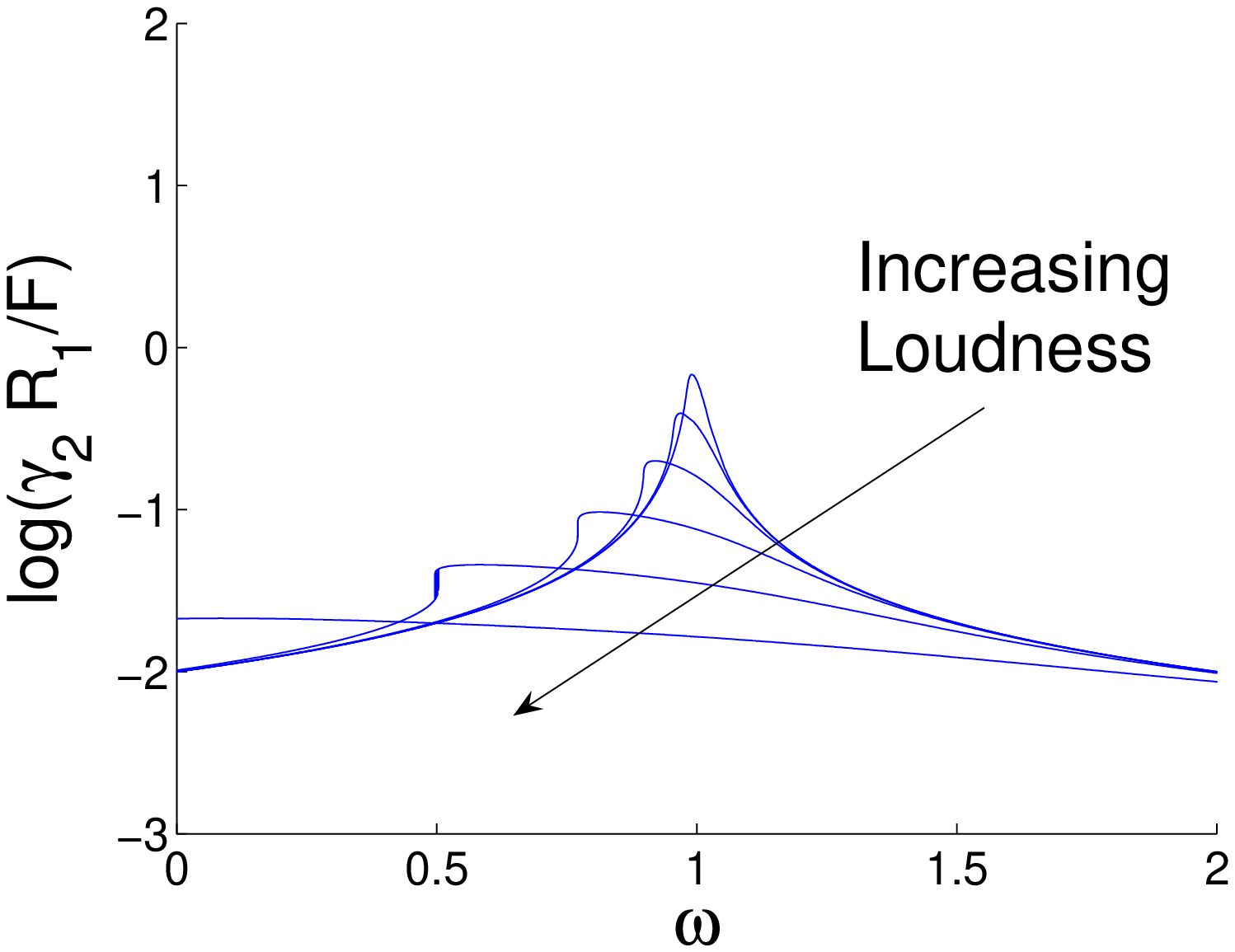}}\resizebox{2 in}{!}{\includegraphics{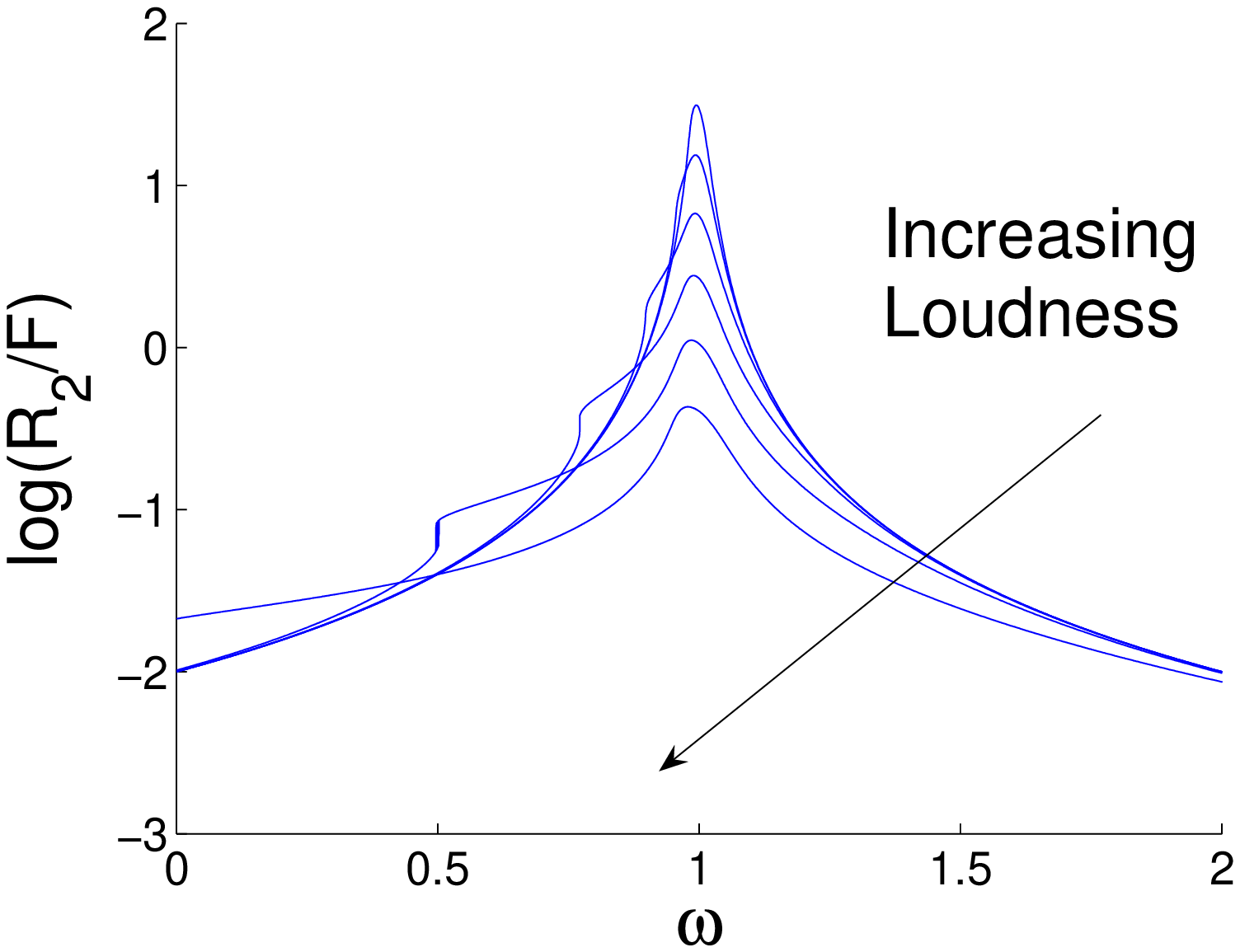}}\resizebox{2 in}{!}{\includegraphics{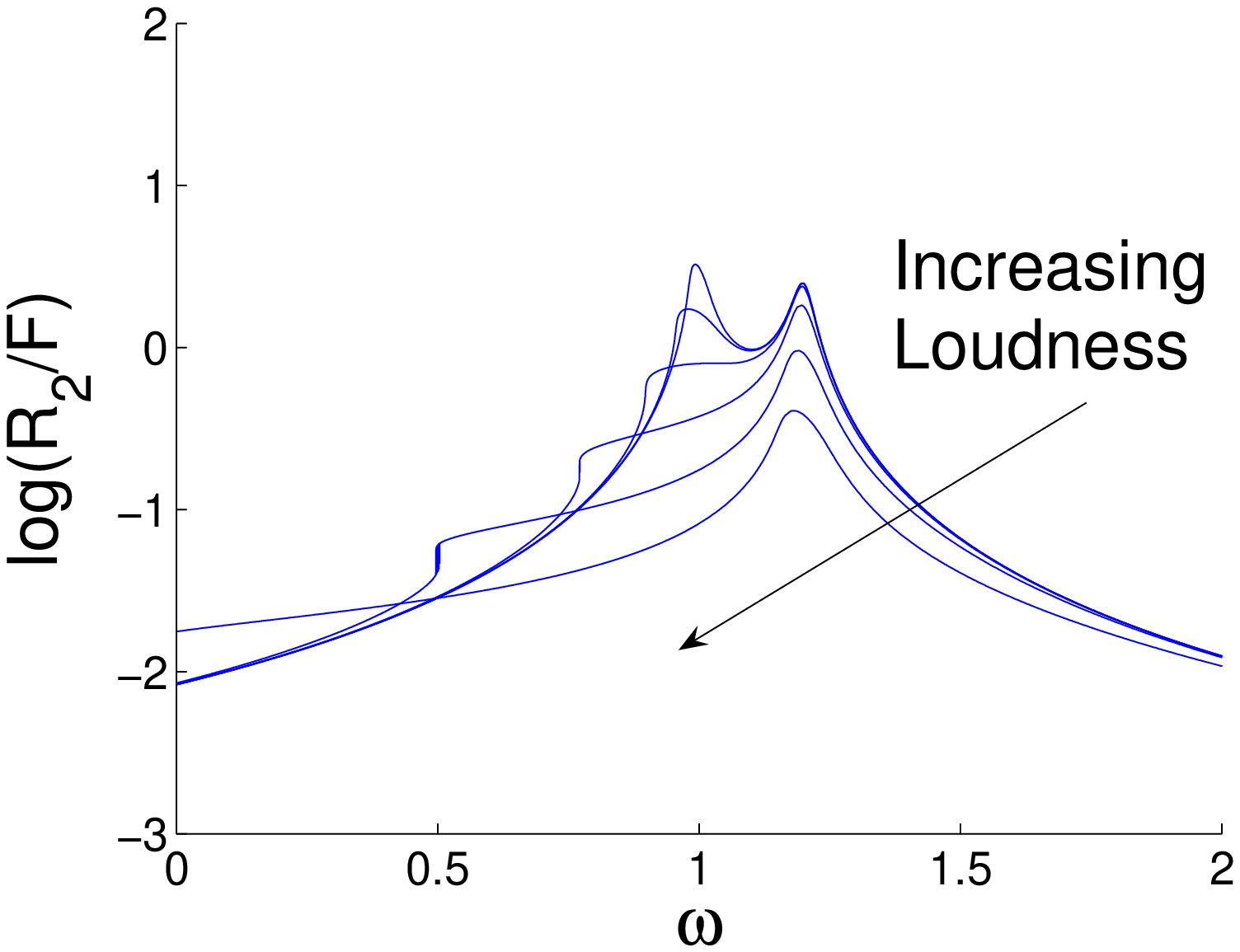}}}
\hspace{.2 in}(a) \hspace{1.8 in} (b) \hspace{1.8 in} (c)
\caption{Amplification {\it vs.} frequency plots
from (\ref{eq:R1}-\ref{eq:poly}).
Amplification was taken to be $\gamma_2 R_{1}/F$ in the single tuning
mechanism case (a), and $R_2/F$ in the double-tuning,
uni-directionally coupled cases (b-c).  Each curve represents a constant
forcing amplitude taken from the set
$F=10^{-3},10^{-2.5},10^{-2.0}...10^{-.5}$ (a) $a_1=-.01$, $b_1=1$,
$c_1=-1$, and $d_1=-2$, $\gamma_2=.01$, $\psi_2=0.64\pi$.  (b)
$a_1=-.01$, $a_2=-.02$, $b_1=b_2=1$,
$c_1=c_2=-1$, $d_1=-2$, $d_2=-1.1$, $\gamma_1=0$, $\gamma_2=0.01$, 
$\psi_2=0.64\pi$.  (c) $a_1=-.01$,
$a_2=-.02$, $b_1=1$, $b_2=1.2$, $c_1=c_2=-1$, $d_1=-2$, $d_2=-1.1$,
$\gamma_1=0$, $\gamma_2=0.01$, $\psi_2=0.64\pi$.  }
\label{fig:singledouble}
\end{figure}

We expect that if each independent amplification mechanism is
well-tuned to the forcing frequency ($|b_j-\omega|\ll 1$) and 
the damping is small the coupled system will have a greater response 
than a system with only a single tuning mechanism. From~(\ref{eq:alphacoeffs}) we see that the leading
coefficients $\alpha_j$ for $j=1,\ldots, 7$ in~(\ref{eq:poly}) may
decrease in magnitude, compared to the highest order coefficient
$\alpha_9$, as the forcing frequency $\omega$ approaches the natural
frequencies of the tuning mechanisms,
$b_j$. Figure~\ref{fig:singledouble} demonstrates that the
amplification, $\frac{R_2}{F}$, is greatest near the resonant
frequency. It also makes a comparison between the tuning curves for
the coupled system and the system with one amplification mechanism
suppressed. For the latter case we solve~(\ref{eq:m1}) for
$R_1(\omega)$ with $R_2=0$, and then plot $\gamma_2 \frac{R_1}{F}$ as
a function of frequency, for different values of $F$; the coupling
$\gamma_2$ is included so that the same relationship between the hair
bundle displacement and the magnitude of the transduction current is
assumed in both the single and the coupled tuning models.  Each curve
in the diagram shows the variation in response with forcing frequency,
holding the signal amplitude $F$
constant. Figure~\ref{fig:singledouble}(a) shows that the broadest
tuning curve occurs for the loudest sounds.  As the magnitude of the
sound decreases, several effects occur: the amplification increases,
the frequency tuning becomes sharper, and because $d_j \neq 0$ the
preferred frequency shifts as the magnitude of the forcing increases.
We used $d_j<0$ for all of the plots, so the preferred frequencies of
the oscillation mechanisms decrease as their amplitude increases. This
phenomenon in turn leads to a region of bistability for driving
frequencies $\omega<b_j$ as described in Section~\ref{analysis}. This
bistability is the source of the prominent shoulder in the tuning curves
that appears for $\omega<1$.  A comparison of the height of the peaks
in Figure~\ref{fig:singledouble}(a) and (b), reveals that
amplification of on-resonance forcing is enhanced in the coupled
system. Also, the frequency tuning curves are sharper and display a
smaller shift in frequency with changing forcing amplitude in the
coupled system.  Each of these properties would be a potential
advantage of the coupled system. If the preferred frequencies of the
two mechanisms are not sufficiently well--tuned to each other, then
the single resonance peak in Fig.~\ref{fig:singledouble}(b) may split
into two peaks as seen in Fig.~\ref{fig:singledouble}(c).

The algebraic relationship between the magnitude of the forcing and
the magnitude of the response of the electrical resonance mechanism
given in (\ref{eq:poly}) allows for experimentally testable
predictions to be made.  In an idealized situation, for which each
mechanism is perfectly tuned to the forcing frequency ($b_1= b_2
=\omega$) and situated at the Hopf bifurcation ($a_1=a_2=0$),
equation~(\ref{eq:poly}) indicates that the response of the electrical
resonance mechanism is proportional to $F^{\frac{1}{9}}$.  The
exponent, $\delta$, of this response-versus-forcing relationship, $R_2
\propto F^{\delta}$, provides a measure of the quality of the
amplification; small amplitude signals are amplified to a greater
extent for smaller values of $\delta$.  For a system employing only a
single tuning mechanism, critically tuned to the bifurcation point and
forcing frequency, the expected scaling is $R \propto F^{\frac{1}{3}}$
\cite{EOCHM00,CDJP00}.  The smaller exponent of $\delta=1/9$,
associated with the well-tuned coupled model, can provide more
powerful amplification ($R_2/F$) than a system with an isolated tuning
mechanism. In more realistic situations for which some damping and/or
detuning is present ($a_j<0$, $b_j \neq \omega$), equation
(\ref{eq:poly}) predicts a transition from a regime in which $R_2
\propto F$ for smaller forcings and a regime in which $R_2 \propto
F^{1/9}$ for sufficiently large signals.  By measuring the response of
the electrical resonance mechanism for different amplitudes of signal,
it is possible to estimate $\delta$, although, depending upon which
portion of the response-versus-forcing curve is sampled, different
estimates for $\delta$ might be obtained experimentally.

In comparing the scaling predictions of our model to experiment, we
turn to auditory nerve data.  Changes in the membrane potential of the
hair cell body result in the release of neurotransmitters at the hair
cell-auditory nerve synapse.  Larger depolarizations result in larger
amounts of neurotransmitter release and thus a faster firing rate in
the auditory nerve.  There is biological evidence that, even for
nonresonant stimuli, the firing rate at the auditory nerve is a
nonlinear function of the sound stimulus. However, this nonlinearity,
which occurs either due to synaptic effects or the relationship
between hair bundle displacement and DC receptor potential, is
factored out during the data analysis, allowing an estimate of
$\delta$ to be made (see discussion in \cite{YWR90,YMK00}).

Numerous experiments have been performed in order to estimate the
exponent, $\delta$, of the response-versus-forcing relationship from
experimental auditory nerve recordings.  Table I summarizes some
recent measurements of $\delta$ taken from experiments in both
mammals and nonmammals.  In each case, $\delta$ is commonly measured
to be smaller than 1/3.  In mammals, as in nonmammals, two different
mechanisms have been proposed to explain amplification, the first due
to active motion of outer hair cells \cite{A87,SD88}, and the second
due to the active motion of the hair bundles
\cite{cf85,MH99,BMH96,KCF05}.  While our results do not directly apply to
mammalian data, as the coupling of the two active elements may be more
complicated \cite{KECF06}, it is noteworthy that compression
estimates in mammals are similar to those in nonmammals.

\vspace{.1 in}
\noindent \begin{tabular}{|l|}
\hline
Table I. Experimental estimates of $\delta$ \\
\hline
Nonmammals: \\
\hline
\textbf{Owls:} between 0.05 and 0.55, with the majority  
of data points lying between 0.1 and 0.3 \cite{KY99}  \\
\textbf{Pigeons:} between 0.22 and 0.6 \cite{RHK95}  \\                                          
\hline
Mammals: \\
\hline
\textbf{Guinea-pigs:} between 0.2 and 0.25 \cite{YWR90}. \\
\textbf{Guinea-pigs:} approximately 0.6 for two low- 
frequency (1.8 and 2.7 kHz.) fibers, \\ and approximately 0.1 for 
medium- (5.5-6.3 kHz) and high-frequency 
(20.5-23 kHz) fibers.\\  For fibers tuned above 
4 kHz, the mean exponent was 0.13 with a standard 
deviation of 0.04  \cite{CY94}. \\
\textbf{Chinchilla:} direct basilar membrane measurements 
yield $\delta$ values between 0.2 and 0.7 \cite{RRRNR97}. \\  
\hline
\end{tabular}
\vspace{.1 in}

The observation of $\delta$ values smaller than 1/3 is interesting in
that it indicates that the auditory system achieves greater
compressions than would be expected from a system with a single tuning
mechanism associated with a Hopf bifurcation.  Experimental
measurements of the forcing-versus-displacement relationship for
individual hair bundles satisfy the $R \propto F^{1/3}$ scaling law
expected for a system tuned near a generic Hopf bifurcation
\cite{EOCHM00,CDJP00}.  Because effects at the synapse are removed
during the data analysis, any additional compression must occur due to
the interaction with the electrical resonance mechanism.

Some models have explained this increased compression by assuming
within their model, that the leading nonlinear terms are higher than
cubic (e.g.\cite{Y90}).  We propose that a more physically motivated
way of achieving higher order compression would be through the
coupling of two systems tuned near a Hopf bifurcation.  With the
exception of three data points, all $\delta$-values in K\"{o}ppl and
Yates' owl data are greater than .1, with the majority of measurements
lying between .1 and .3 \cite{Y90}.  So their data is not 
inconsistent with what would be expected from a coupled system, which
at best produces a $\delta$-value of 1/9.  If it is the case that the
observed enhanced amplification occurs due to coupling between the two
tuning mechanisms, this is an interesting result, because our analysis
indicates that the tuning mechanisms must maintain themselves close to
the bifurcation point and close to the same frequency for maximum
amplification to be observed.  It would be interesting both from a
biological and mathematical perspective to understand how such fine
tuning is achieved.  Some studies have suggested that stochastic 
effects may help the system adjust itself to the bifurcation point \cite{NMJ04,B05}.
Another has suggested that with certain assumptions about the evolution
of the bifurcation parameter, self-tuning occurs automatically \cite{MS03}.
Biologically, it has been suggested that adjustments in the
tension of the hair bundle due to an actin myosin mechanism
may act to keep the hair bundle properly tuned \cite{VD03,MBCH03}.

Another  prediction of the model is that the response of the
electrical resonance mechanism may be hysteretic for sufficiently large
frequency detuning.  It is worth noting that this feature arises as a
direct result of shift in the natural frequency of the tuning
mechanisms with increasing response, a feature that was not taken
into account in previous models.  It would be interesting if such 
bistability could be observed in experimental auditory nerve data.

\acknowledgments

K.A.M. is grateful for fellowship support through NSF-IGERT grant
DGE-9987577  and NSF-RTG grant 0354259.  M.S. acknowledges support through NSF Grant DMS-0309667.

\section{Appendix I}
\label{sec:appendix}
\textbf{{Model Proposed by Hudspeth and Lewis}}

The following seven-dimensional model of a bullfrog saccular hair cell
was proposed by Hudspeth and Lewis \cite{HL88a,HL88b}.  It 
is a Hodgkin and Huxley type model with three ion channels
included in the model, a voltage-gated calcium ion channel, a
calcium-gated potassium ion channel, and a passive leakage channel.
The passive leakage channel is always open.  The calcium ion channel
opens in response to depolarization. The gating of the potassium ion
channel requires both the binding of calcium to the interior of the
channel and an adequate depolarization.  The first equation models the
rate of change of the membrane potential ($V_m$) due to currents entering the
cell through both the transduction channels, the calcium ion channels,
and the calcium-gated potassium channels.  The other six equations
account for changes in the concentration of internal calcium ($Ca$)
close to the membrane, the fraction of open calcium ion channels
($m$), and the fraction of potassium ion channels in each of three
closed configurations ($C_0, C_1, C_2$) and two open configurations
($O_2,O_3$).

\begin{eqnarray}
\label{eq:hlmodel}
C_m \frac{ dV_{m}}{dt}&=& -g_{Ca} m^3 (V_m-E_{Ca}) \nonumber \\
&&-g_{K(Ca)}(O_2+O_3)(V_m - E_K) \nonumber \\
&& - g_L (V_m - E_L) +I \nonumber\\
\frac{d Ca}{dt}&=&-\frac{U g_{Ca} m^3 (V_m-E_{Ca})}{z F v_{cell} \xi}-K_s Ca \nonumber\\
\frac{dm}{dt}&=& \beta(V_m) (1-m)- \alpha(V_m) m \nonumber\\
\frac{dC_0}{dt}&=& k_{-1} C_1 - k_1 Ca C_0 \\
\frac{dC_1}{dt} &=& k_1 Ca C_0 + k_{-2} C_2 - (k_{-1} + k_2 Ca) C_1 \nonumber\\
\frac{dC_2}{dt}& =& k_2 Ca C_1 + \alpha_{C0} e^{-Vm/V_{aa}} O_2 - (k_{-2}+ \beta_C) C_2 \nonumber\\
\frac{dO_2}{dt}& =& \beta_{C} C_2 + k_{-3} O_3 - (\alpha_{C0} e^{-Vm/V_{aa}} + k_3 Ca) O_2 \nonumber\\
O_3 &=& 1 - C_{0} - C_{1} - C_{2} - O_{2} \nonumber 
\end{eqnarray}

\noindent where

\begin{eqnarray}
\alpha(V_m) &=&\alpha_0 e^{-(V_m+V_0)/V_A} + K_A \nonumber \\
\beta(V_m) &=& \beta_0 e^{(V_m+V_0)/V_B}+K_B \nonumber \\
k_1 &=& \frac{k_{-1}}{K_{10} e^{\frac{\delta_1 z F V_m}{R T}}}  \nonumber \\
k_2 &=& \frac{k_{-2}}{K_{20} e^{\frac{\delta_2 z F V_m}{R T}}} \nonumber \\
k_3 &=& \frac{k_{-3}}{K_{30} e^{\frac{\delta_3 z F V_m}{R T}}} \nonumber
\end{eqnarray}

The amount of current that is injected into the cell, $I$, is used as
the control parameter.  All other parameters are set to the value used
in Hudspeth and Lewis's original paper: $g_{Ca}=4.14\times 10^{-9}$~S,
$E_{Ca}=.1$~V, $g_{K_{Ca}}=16.8\times 10^{-9}$~S, $E_K=-.08$~V,
$E_L=-.03$~V, $g_L=10^{-9}$~S, $C_m=15\times 10^{-12}$~F,
$U=.02$,$z=2$, $F=96485.309$~C/mol, $v_{cell}=1.25\times 10^{-12} L$,
$\xi=3.4\times 10^{-5}$, $K_s=2800$~$s^{-1}$, $\beta_0=.97$~$s^{-1}$,
$V_0=.07$~V, $V_B=.00617$~V, $K_B=940$~$s^{-1}$, $K_A=510$~$s^{-1}$,
$V_A=.00801$~V, $\alpha_0=22800$~$s^{-1}$, $\alpha_{C0}=450$~$s^{-1}$,
$T=295$ K, $R=8.314510$~$\frac{J}{mol-K}$, $\delta_1=.2$, $\delta_2=0$,
$\delta_3=.2$, $\beta_C=1000$~$s^{-1}$, $k_{-1}=300$~$s^{-1}$,
$k_{-2}=5000$~$s^{-1}$, $k_{-3}=1500$~$s^{-1}$, $K_{10}=6\times
10^{-6}$~M, $K_{20}=45\times 10^{-6}$~M, $K_{30}=20\times 10^{-6}$~M,
and $V_{aa}=.033$~V.

\section{Appendix II}
\label{sec:appendix2}

Here we present some details of the reduction of the Hudspeth and
Lewis model (\ref{eq:hlmodel}) to the normal form
(\ref{eq:singleredf}).  Our analysis is valid in a neighborhood of the
equilibrium solution $(V_m^*,Ca^*, m^*,C_0^*,C_1^*,C_2^*, O_2^*)$ for
command currents $I$ sufficiently close to the critical current $I^*$.
In particular, we find that $V_{m}^* \approx -0.04888$ V, $Ca^*
\approx 1.623\times 10^{-5}$, $m^* \approx 0.3115$, $C_{0}^* \approx
0.08429$, $C_{1}^* \approx 0.4919$, $C_{2}^* \approx 0.1774$, $O_{2}^*
\approx 0.08960$, and $O_{3}^* \approx 0.1568$, for $I=I^*\approx
91.3\times 10^{-12}$ A.

We first translate the fixed point to the origin and nondimensionalize
variables as follows: $V_{m}=V_{m}^*(1+X_1)$, $Ca=Ca^*(1+X_2)$,
$m=m^*(1+X_3)$, $C_{0}=C_{0}^*(1+X_4)$, $C_{1}=C_{1}^* (1+X_5)$,
$C_{2}=C_{2}^*(1+X_6)$, $O_{2}=O_{2}^*(1+X_7)$.  Moreover, we let
$I=I^*(1+\Delta I)$, where $\Delta I$ measures a small deviation from
the critical command current associated with the Hopf bifurcation, and
allow for a small periodic forcing through the leakage current
conductance by setting $g_L=g_L^*(1+\Delta g_L(t))$.  Here $\Delta
g_L(t) =\Delta g_L(t+T)$ captures the purely oscillatory part of
$g_L$, while $g_L^*$ is the mean conductance; we set $g_L^*=10^{-9}$
S, as in the original Hudspeth and Lewis model.  The period $T$ is
related to the forcing frequency $\omega$ in the usual fashion,
$T\equiv 2\pi/\omega$. The governing equations, expressed in terms of
the dimensionless vector-valued variable ${\bf
X}=(X_1,X_2,\ldots,X_7)$ and the parameters $\Delta I$ and $\Delta
g_L(t)$, are written
\begin{equation}
  {d{\bf X}\over dt}={\bf H}({\bf X}; \Delta I,\Delta g_L(t))\ .
  \label{eq:g1}
\end{equation}
The Hopf bifurcation occurs at ${\bf X}={\bf 0}$, $\Delta I=0$ for
$\Delta g_L(t)=0$, so ${\bf H}({\bf 0};0,0)={\bf 0}$ and the Jacobian
matrix ${\bf DH}({\bf 0};0,0)$ has a pair of purely imaginary
eigenvalues $\pm i\omega_0,$ $\omega_0\approx 938$ $s^{-1}$, with
associated complex eigenvectors ${\bf U},\overline{\bf U}$. The
remaining eigenvalues $\lambda_1,\ldots,\lambda_5$ (where
$\lambda_5=\overline{\lambda_4}$) all have real parts less than
$-2000$ $s^{-1}$ (see Fig.~\ref{fig:hudeigs}), and are associated with
eigenvectors ${\bf V}_1,\ldots,{\bf V}_5$ (where, again, ${\bf
V}_5=\overline{\bf V}_4$).  Our convention is to normalize all
eigenvectors to 1. Finally, we diagonalize the linearized problem at
the bifurcation point by letting ${\bf X}(t)=z(t){\bf U}+
\overline{z}(t)\overline{\bf U}+y_1(t){\bf V}_1+ y_2(t){\bf
V}_2+\cdots y_5(t){\bf V}_5$, and write the governing equations in
terms of the new variables $z,\overline{z}, {\bf
y}\equiv(y_1,\cdots,y_5)$ as follows:
\begin{eqnarray}
\label{eq:zyeqs}
{dz\over dt}&=&i\omega_0 z+N_z(z,\overline{z},{\bf y};\Delta I,\Delta g_L(t)),
\nonumber\\
{d\overline{z}\over dt}&=&-i\omega_0 \overline{z}+\overline{N}_z
(z,\overline{z},{\bf y};\Delta I,\Delta g_L(t)),\\
{d{\bf y}\over dt}&=& 
\Lambda{\bf y }+{\bf N}_{\bf y}(z,\overline{z},
{\bf y};\Delta I,\Delta g_L(t))\ .
\nonumber
\end{eqnarray}
Here $\Lambda$ is a diagonal matrix, with eigenvalues
$\lambda_1,\ldots, \lambda_5$ on the diagonal, and $N_z$ and ${\bf
N}_{\bf y}$ contain the nonlinear terms in $z,\overline{z}$ and ${\bf
y}$, as well as all terms involving the parameters $\Delta I$ and
$\Delta g_L(t)$.

We now use perturbation theory to derive the normal form of the
bifurcation problem. Toward this end we introduce a small book-keeping
parameter $\epsilon$ ($|\epsilon|\ll 1$) that is a measure of
proximity to the Hopf bifurcation. Specifically, we let $\Delta
I=\epsilon^2\mu$ and seek small amplitude solutions of the form
\begin{eqnarray}
\label{eq:epsexp}
z(t)&=&\epsilon z_1(t,T)+\epsilon^2 z_2(t,T)+\epsilon^3 z_3(t,T)+\cdots,\\
{\bf y}(t)&=&\epsilon^2 {\bf y}_2(t,T)+\epsilon^3 {\bf y}_3(t,T)+\cdots,
\nonumber
\end{eqnarray}
where $T=\epsilon^2 t$ is a slow time variable that captures the slow
decay to the oscillatory solutions associated with a Hopf bifurcation.
Finally, we make some additional assumptions about the magnitude and
frequency of the applied periodic forcing by letting $\Delta
g_L(t)=\epsilon^3 f(t)$ and $\omega_0=\omega+\epsilon^2\hat\omega$,
{\it i.e.} we consider small nearly-resonant periodic forcing.

Inserting these ansatz in equation~(\ref{eq:zyeqs}), and expanding in
powers of $\epsilon$, we recover at ${\cal O}(\epsilon)$
\begin{equation}
{\partial z_1\over \partial t}=i\omega z_1,
\end{equation}
with solution
\begin{equation}
z_1=\hat A(T)e^{i\omega t}\ .
\label{eq:z1sol}
\end{equation}
Here the complex amplitude $\hat A(T)$ satisfies an equation to be
determined at higher order.

At ${\cal O}(\epsilon^2)$, we find
\begin{equation}
{\partial z_2\over \partial t}=
i\omega z_2+\alpha_z\mu+\beta_z z_1^2+\gamma_z\overline z_1^2+\delta_z|z_1|^2,
\label{eq:z2dot}
\end{equation}
where $\alpha_z\approx -756+52i$, $\beta_z\approx -86+2i$,
$\gamma_z\approx 20-20i$, $\delta_z\approx -80-64i$, as well as similar
equations for the fast-time evolution of ${\bf y}_2$. The general solution 
of~(\ref{eq:z2dot}) is
\begin{equation}
z_2(t,T) = \frac{i(\alpha_z\mu+\delta_z|\hat A|^2)}{\omega}
-\frac{i\beta_z\hat A^2}{\omega}e^{2i\omega t}
+\frac{i\gamma_z{\overline{\hat A}}^2}{3\omega}e^{-2i\omega t}+\hat B(T)e^{i\omega t}\ .
\label{eq:z2sol} \end{equation}
Here $\hat B(T)$ is arbitrary and we set it to zero in the remainder
of our computations, since, without loss of generality, it may be
absorbed by the ${\cal O}(\epsilon)$ solution~(\ref{eq:z1sol}). The
equations for each of the components of ${\bf y}_2$, which have a
similar structure to~(\ref{eq:z2dot}), yield solutions of the same
form as~(\ref{eq:z2sol}), although with $\hat B(T)e^{i\omega t}$
replaced by rapidly decaying solutions $C_j(T)e^{\lambda_j t}$ of
their associated homogeneous problems.

Finally, at ${\cal O}(\epsilon^3)$, we find
\begin{equation}
{\partial z_3\over \partial t} = i\omega z_3+\Bigl(i\hat\omega
\hat A-{\partial \hat A\over \partial T} +(45+97i)\mu
\hat A 
-(46+49i)|\hat A|^2\hat A -(156-11i)\hat f_1\Bigr)e^{i\omega t}+\cdots.  
\end{equation}
Here the ellipsis indicates additional terms proportional to
$e^{-i\omega t}$ and $e^{\pm 3i\omega t}$, and $\hat f_1$ is the
coefficient of the $e^{i\omega t}$ term in the Fourier expansion of
the periodic forcing function $f(t)$. We have written explicitly only
those terms on the right-hand-side that are resonant with the solution
of the linear homogeneous problem. In order for our perturbation
expansion to remain valid, this resonant forcing term, which leads to
unbounded growth, must vanish. In this fashion we obtain the following
evolution equation for $\hat A(T)$:
\begin{equation}
{\partial \hat A\over \partial T}
 = i\hat\omega
\hat A +(45+97i)\mu
\hat A-(46+49i)|\hat A|^2\hat A -(156-11i)\hat f_1\ .
\end{equation}
Finally, we re-write the equation in the normal form~(\ref{eq:singleredf}) 
by letting 
\begin{equation}
A(t)=\epsilon \hat A(\epsilon^2 t)e^{i\omega t+i\varphi}\ , 
\end{equation}
where the phase $\varphi$ is specified below.
We then find
\begin{equation}
{dA\over dt}=(a+ib)A+(c+id)|A|^2A+Fe^{i\omega t},
\label{eq:finalresult}
\end{equation}
where
\begin{eqnarray}
a+ib &=& 45\Delta I+i(\omega_0+97\Delta I),\quad  \nonumber \\
c+id &=&-(46+49i) ,\quad \nonumber \\
F &=& -(156-11i)\widehat{\Delta g}_{L,1}e^{i\varphi}\ .
\end{eqnarray}
Here $\widehat{\Delta g}_{L,1}$ is the coefficient of the
$e^{i\omega t}$ term in the Fourier expansion of $\Delta g_L$, and we choose
$\varphi$ so that $F$ is real and positive, {\it i.e.} so that
\begin{equation}
F=|(156-11i)\widehat{\Delta g}_{L,1}|\ .
\end{equation}

Next, we will consider how two systems tuned near a Hopf bifurcation
would interact.  Under the assumption that the coupling between the
two systems is weak and linear, it's straightforward to show that the
coupled system can be described by the following set of coupled
amplitude equations.

\begin{equation}
  \frac{dA_1}{dt}= (a_1 + b_1 i) A_1 +(c_1 + d_1 i) |A_1|^2 A_1 + \gamma_1 e^{i \psi_1} A_2+ F e^{i \omega t},
\label{eq:coupledamp1}
\end{equation}

\begin{equation}
  \frac{dA_2}{dt}=(a_2 + b_2 i) A_2 +(c_2 + d_2 i) |A_2|^2 A_2 + \gamma_2 e^{i \psi_2} A_2,
\label{eq:coupledamp2}
\end{equation}

\noindent where the coefficients of the coupling terms, $\gamma_1$ and
$\gamma_2$, are taken to be of order $\epsilon^2$ indicating weak
coupling between the two systems.  Additionally, for coupling between
the two systems to occur, it is necessary to assume that the resonance
frequency of each system is within $\epsilon^2$ of the forcing
frequency $\omega$.  As with the previous amplitude equations, the
resonant forcing amplitude, $F$, is assumed to be small on the order
of $\epsilon^3$ and each system is assumed to be tuned sufficiently
close to the Hopf bifurcation, requiring that both $a_1$ and $a_2$ be
of order $\epsilon^2$.


\begin{thebibliography}{41}
\expandafter\ifx\csname natexlab\endcsname\relax\def\natexlab#1{#1}\fi
\expandafter\ifx\csname bibnamefont\endcsname\relax
  \def\bibnamefont#1{#1}\fi
\expandafter\ifx\csname bibfnamefont\endcsname\relax
  \def\bibfnamefont#1{#1}\fi
\expandafter\ifx\csname citenamefont\endcsname\relax
  \def\citenamefont#1{#1}\fi
\expandafter\ifx\csname url\endcsname\relax
  \def\url#1{\texttt{#1}}\fi
\expandafter\ifx\csname urlprefix\endcsname\relax\def\urlprefix{URL }\fi
\providecommand{\bibinfo}[2]{#2}
\providecommand{\eprint}[2][]{\url{#2}}

\bibitem[{\citenamefont{Robles and Ruggero}(2001)}]{RR01}
\bibinfo{author}{\bibfnamefont{L.}~\bibnamefont{Robles}} \bibnamefont{and}
  \bibinfo{author}{\bibfnamefont{M.~A.} \bibnamefont{Ruggero}},
  \bibinfo{journal}{Physiol. Rev.} \textbf{\bibinfo{volume}{81(3)}},
  \bibinfo{pages}{1305} (\bibinfo{year}{2001}).

\bibitem[{\citenamefont{Manley}(2001)}]{M01}
\bibinfo{author}{\bibfnamefont{G.~A.} \bibnamefont{Manley}},
  \bibinfo{journal}{J. Neurophysiol.} \textbf{\bibinfo{volume}{86(2)}},
  \bibinfo{pages}{541} (\bibinfo{year}{2001}).

\bibitem[{\citenamefont{Fettiplace et~al.}(2001)\citenamefont{Fettiplace,
  Ricci, and Hackney}}]{FRH01}
\bibinfo{author}{\bibfnamefont{R.}~\bibnamefont{Fettiplace}},
  \bibinfo{author}{\bibfnamefont{A.~J.} \bibnamefont{Ricci}}, \bibnamefont{and}
  \bibinfo{author}{\bibfnamefont{C.~M.} \bibnamefont{Hackney}},
  \bibinfo{journal}{Trends in Neurosciences} \textbf{\bibinfo{volume}{24(3)}},
  \bibinfo{pages}{169} (\bibinfo{year}{2001}).

\bibitem[{\citenamefont{Kandel et~al.}(2000)\citenamefont{Kandel, Schwartz, and
  Jessell}}]{Kandel}
\bibinfo{author}{\bibfnamefont{E.~R.} \bibnamefont{Kandel}},
  \bibinfo{author}{\bibfnamefont{J.~H.} \bibnamefont{Schwartz}},
  \bibnamefont{and} \bibinfo{author}{\bibfnamefont{T.~M.}
  \bibnamefont{Jessell}}, \emph{\bibinfo{title}{Principles of Neural Science,
  Fourth Edition}} (\bibinfo{publisher}{McGraw-Hill}, \bibinfo{address}{New
  York}, \bibinfo{year}{2000}).

\bibitem[{\citenamefont{Hudspeth}(1989)}]{H89}
\bibinfo{author}{\bibfnamefont{A.~J.} \bibnamefont{Hudspeth}},
  \bibinfo{journal}{Nature} \textbf{\bibinfo{volume}{341(6241)}},
  \bibinfo{pages}{397} (\bibinfo{year}{1989}).

\bibitem[{\citenamefont{Martin and Hudspeth}(1999)}]{MH99}
\bibinfo{author}{\bibfnamefont{P.}~\bibnamefont{Martin}} \bibnamefont{and}
  \bibinfo{author}{\bibfnamefont{A.~J.} \bibnamefont{Hudspeth}},
  \bibinfo{journal}{P. Natl. Acad. Sci.} \textbf{\bibinfo{volume}{96(25)}},
  \bibinfo{pages}{14306} (\bibinfo{year}{1999}).

\bibitem[{\citenamefont{Fettiplace and Crawford}(1978)}]{FC78}
\bibinfo{author}{\bibfnamefont{R.}~\bibnamefont{Fettiplace}} \bibnamefont{and}
  \bibinfo{author}{\bibfnamefont{A.~C.} \bibnamefont{Crawford}},
  \bibinfo{journal}{P. Roy. Soc. Lond. B} \textbf{\bibinfo{volume}{203}},
  \bibinfo{pages}{209} (\bibinfo{year}{1978}).

\bibitem[{\citenamefont{Choe et~al.}(1998)\citenamefont{Choe, Magnasco, and
  Hudspeth}}]{CMH98}
\bibinfo{author}{\bibfnamefont{Y.}~\bibnamefont{Choe}},
  \bibinfo{author}{\bibfnamefont{M.~O.} \bibnamefont{Magnasco}},
  \bibnamefont{and} \bibinfo{author}{\bibfnamefont{A.~J.}
  \bibnamefont{Hudspeth}}, \bibinfo{journal}{P. Natl. Acad. Sci.}
  \textbf{\bibinfo{volume}{95(26)}}, \bibinfo{pages}{15321}
  (\bibinfo{year}{1998}).

\bibitem[{\citenamefont{Ospeck et~al.}(2001)\citenamefont{Ospeck, Egu\'{i}luz,
  and Magnasco}}]{OEM01}
\bibinfo{author}{\bibfnamefont{M.}~\bibnamefont{Ospeck}},
  \bibinfo{author}{\bibfnamefont{V.~M.} \bibnamefont{Egu\'{i}luz}},
  \bibnamefont{and} \bibinfo{author}{\bibfnamefont{M.~O.}
  \bibnamefont{Magnasco}}, \bibinfo{journal}{Biophys. J.}
  \textbf{\bibinfo{volume}{80(6)}}, \bibinfo{pages}{2597}
  (\bibinfo{year}{2001}).

\bibitem[{\citenamefont{Egu\'{i}luz et~al.}(2000)\citenamefont{Egu\'{i}luz,
  Ospeck, Choe, Hudspeth, and Magnasco}}]{EOCHM00}
\bibinfo{author}{\bibfnamefont{V.~M.} \bibnamefont{Egu\'{i}luz}},
  \bibinfo{author}{\bibfnamefont{M.}~\bibnamefont{Ospeck}},
  \bibinfo{author}{\bibfnamefont{Y.}~\bibnamefont{Choe}},
  \bibinfo{author}{\bibfnamefont{A.~J.} \bibnamefont{Hudspeth}},
  \bibnamefont{and} \bibinfo{author}{\bibfnamefont{M.~O.}
  \bibnamefont{Magnasco}}, \bibinfo{journal}{Phys. Rev. Lett.}
  \textbf{\bibinfo{volume}{84(22)}}, \bibinfo{pages}{5232}
  (\bibinfo{year}{2000}).

\bibitem[{\citenamefont{Camalet et~al.}(2000)\citenamefont{Camalet, Duke,
  J\"{u}licher, and Prost}}]{CDJP00}
\bibinfo{author}{\bibfnamefont{S.}~\bibnamefont{Camalet}},
  \bibinfo{author}{\bibfnamefont{T.}~\bibnamefont{Duke}},
  \bibinfo{author}{\bibfnamefont{F.}~\bibnamefont{J\"{u}licher}},
  \bibnamefont{and} \bibinfo{author}{\bibfnamefont{J.}~\bibnamefont{Prost}},
  \bibinfo{journal}{P. Natl. Acad. Sci.} \textbf{\bibinfo{volume}{97(7)}},
  \bibinfo{pages}{3183} (\bibinfo{year}{2000}).

\bibitem[{\citenamefont{Hudspeth and Lewis}(1988{\natexlab{a}})}]{HL88a}
\bibinfo{author}{\bibfnamefont{A.~J.} \bibnamefont{Hudspeth}} \bibnamefont{and}
  \bibinfo{author}{\bibfnamefont{R.~S.} \bibnamefont{Lewis}},
  \bibinfo{journal}{J. Physiol. (Lond.)} \textbf{\bibinfo{volume}{400}},
  \bibinfo{pages}{237} (\bibinfo{year}{1988}{\natexlab{a}}).

\bibitem[{\citenamefont{Hudspeth and Lewis}(1988{\natexlab{b}})}]{HL88b}
\bibinfo{author}{\bibfnamefont{A.~J.} \bibnamefont{Hudspeth}} \bibnamefont{and}
  \bibinfo{author}{\bibfnamefont{R.~S.} \bibnamefont{Lewis}},
  \bibinfo{journal}{J. Physiol. (Lond.)} \textbf{\bibinfo{volume}{400}},
  \bibinfo{pages}{275} (\bibinfo{year}{1988}{\natexlab{b}}).

\bibitem[{\citenamefont{Wiggins}(1990)}]{W1990}
\bibinfo{author}{\bibfnamefont{S.}~\bibnamefont{Wiggins}},
  \emph{\bibinfo{title}{Introduction to Applied Nonlinear Dynamical Systems and
  Chaos}}, vol.~\bibinfo{volume}{2} of \emph{\bibinfo{series}{Texts in Applied
  Mathematics}} (\bibinfo{publisher}{Springer-Verlag}, \bibinfo{address}{New
  York}, \bibinfo{year}{1990}).

\bibitem[{\citenamefont{Cheung and Corey}(2006)}]{CC06}
\bibinfo{author}{\bibfnamefont{E.~L.~M.} \bibnamefont{Cheung}}
  \bibnamefont{and} \bibinfo{author}{\bibfnamefont{D.~P.} \bibnamefont{Corey}},
  \bibinfo{journal}{Biophys. J.} \textbf{\bibinfo{volume}{90}},
  \bibinfo{pages}{124} (\bibinfo{year}{2006}).

\bibitem[{\citenamefont{Martin and Hudspeth}(2001)}]{MH01}
\bibinfo{author}{\bibfnamefont{P.}~\bibnamefont{Martin}} \bibnamefont{and}
  \bibinfo{author}{\bibfnamefont{A.~J.} \bibnamefont{Hudspeth}},
  \bibinfo{journal}{P. Natl. Acad. Sci.} \textbf{\bibinfo{volume}{98(25)}},
  \bibinfo{pages}{14386} (\bibinfo{year}{2001}).

\bibitem[{\citenamefont{Hudspeth and Corey}(1977)}]{HC77}
\bibinfo{author}{\bibfnamefont{A.~J.} \bibnamefont{Hudspeth}} \bibnamefont{and}
  \bibinfo{author}{\bibfnamefont{D.~P.} \bibnamefont{Corey}},
  \bibinfo{journal}{P. Natl. Acad. Sci.} \textbf{\bibinfo{volume}{74(6)}},
  \bibinfo{pages}{2407} (\bibinfo{year}{1977}).

\bibitem[{\citenamefont{Assad and Corey}(1992)}]{AC92}
\bibinfo{author}{\bibfnamefont{J.~A.} \bibnamefont{Assad}} \bibnamefont{and}
  \bibinfo{author}{\bibfnamefont{D.~P.} \bibnamefont{Corey}},
  \bibinfo{journal}{J. Neurosci.} \textbf{\bibinfo{volume}{12(9)}},
  \bibinfo{pages}{3291} (\bibinfo{year}{1992}).

\bibitem[{\citenamefont{Denk and Webb}(1992)}]{DW92}
\bibinfo{author}{\bibfnamefont{W.}~\bibnamefont{Denk}} \bibnamefont{and}
  \bibinfo{author}{\bibfnamefont{W.~W.} \bibnamefont{Webb}},
  \bibinfo{journal}{Hearing Res.} \textbf{\bibinfo{volume}{60(1)}},
  \bibinfo{pages}{89} (\bibinfo{year}{1992}).

\bibitem[{\citenamefont{Ricci et~al.}(2000)\citenamefont{Ricci, Crawford, and
  Fettiplace}}]{RCF00}
\bibinfo{author}{\bibfnamefont{A.~J.} \bibnamefont{Ricci}},
  \bibinfo{author}{\bibfnamefont{A.~C.} \bibnamefont{Crawford}},
  \bibnamefont{and}
  \bibinfo{author}{\bibfnamefont{R.}~\bibnamefont{Fettiplace}},
  \bibinfo{journal}{J. Neurosci.} \textbf{\bibinfo{volume}{20(19)}},
  \bibinfo{pages}{7131} (\bibinfo{year}{2000}).

\bibitem[{\citenamefont{Ricci et~al.}(2002)\citenamefont{Ricci, Crawford, and
  Fettiplace}}]{RCF02}
\bibinfo{author}{\bibfnamefont{A.~J.} \bibnamefont{Ricci}},
  \bibinfo{author}{\bibfnamefont{A.~C.} \bibnamefont{Crawford}},
  \bibnamefont{and}
  \bibinfo{author}{\bibfnamefont{R.}~\bibnamefont{Fettiplace}},
  \bibinfo{journal}{J. Neurosci.} \textbf{\bibinfo{volume}{22(1)}},
  \bibinfo{pages}{44} (\bibinfo{year}{2002}).

\bibitem[{\citenamefont{Bozovic and Hudspeth}(2003)}]{BH03}
\bibinfo{author}{\bibfnamefont{D.}~\bibnamefont{Bozovic}} \bibnamefont{and}
  \bibinfo{author}{\bibfnamefont{A.~J.} \bibnamefont{Hudspeth}},
  \bibinfo{journal}{P. Natl. Acad. Sci.} \textbf{\bibinfo{volume}{100(3)}},
  \bibinfo{pages}{958} (\bibinfo{year}{2003}).

\bibitem[{\citenamefont{R\"{u}sch and Thurm}(1990)}]{RT90}
\bibinfo{author}{\bibfnamefont{A.}~\bibnamefont{R\"{u}sch}} \bibnamefont{and}
  \bibinfo{author}{\bibfnamefont{U.}~\bibnamefont{Thurm}},
  \bibinfo{journal}{Hearing Res.} \textbf{\bibinfo{volume}{48(3)}},
  \bibinfo{pages}{247} (\bibinfo{year}{1990}).

\bibitem[{\citenamefont{Martin et~al.}(2003)\citenamefont{Martin, Bozovic,
  Choe, and Hudspeth}}]{MBCH03}
\bibinfo{author}{\bibfnamefont{P.}~\bibnamefont{Martin}},
  \bibinfo{author}{\bibfnamefont{D.}~\bibnamefont{Bozovic}},
  \bibinfo{author}{\bibfnamefont{Y.}~\bibnamefont{Choe}}, \bibnamefont{and}
  \bibinfo{author}{\bibfnamefont{A.~J.} \bibnamefont{Hudspeth}},
  \bibinfo{journal}{J. Neurosci.} \textbf{\bibinfo{volume}{23(11)}},
  \bibinfo{pages}{4533} (\bibinfo{year}{2003}).

\bibitem[{\citenamefont{Yates et~al.}(1990)\citenamefont{Yates, Winter, and
  Robertson}}]{YWR90}
\bibinfo{author}{\bibfnamefont{G.~K.} \bibnamefont{Yates}},
  \bibinfo{author}{\bibfnamefont{I.~M.} \bibnamefont{Winter}},
  \bibnamefont{and}
  \bibinfo{author}{\bibfnamefont{D.}~\bibnamefont{Robertson}},
  \bibinfo{journal}{Hearing Res.} \textbf{\bibinfo{volume}{45(3)}},
  \bibinfo{pages}{203} (\bibinfo{year}{1990}).

\bibitem[{\citenamefont{Yates et~al.}(2000)\citenamefont{Yates, Manley, and
  K\"{o}ppl}}]{YMK00}
\bibinfo{author}{\bibfnamefont{G.~K.} \bibnamefont{Yates}},
  \bibinfo{author}{\bibfnamefont{G.~A.} \bibnamefont{Manley}},
  \bibnamefont{and}
  \bibinfo{author}{\bibfnamefont{C.}~\bibnamefont{K\"{o}ppl}},
  \bibinfo{journal}{J. Acoust. Soc. Am.} \textbf{\bibinfo{volume}{107(4)}},
  \bibinfo{pages}{2143} (\bibinfo{year}{2000}).

\bibitem[{\citenamefont{Ashmore}(1987)}]{A87}
\bibinfo{author}{\bibfnamefont{J.~F.} \bibnamefont{Ashmore}},
  \bibinfo{journal}{J. Physiol. (Lond.)} \textbf{\bibinfo{volume}{388}},
  \bibinfo{pages}{323} (\bibinfo{year}{1987}).

\bibitem[{\citenamefont{Santos-Sacchi and Dilger}(1988)}]{SD88}
\bibinfo{author}{\bibfnamefont{J.}~\bibnamefont{Santos-Sacchi}}
  \bibnamefont{and} \bibinfo{author}{\bibfnamefont{J.~P.}
  \bibnamefont{Dilger}}, \bibinfo{journal}{Hearing Res.}
  \textbf{\bibinfo{volume}{35(2-3)}}, \bibinfo{pages}{143}
  (\bibinfo{year}{1988}).

\bibitem[{\citenamefont{Crawford and Fettiplace}(1985)}]{cf85}
\bibinfo{author}{\bibfnamefont{A.~C.} \bibnamefont{Crawford}} \bibnamefont{and}
  \bibinfo{author}{\bibfnamefont{R.}~\bibnamefont{Fettiplace}},
  \bibinfo{journal}{J. Physiol. (Lond.)} \textbf{\bibinfo{volume}{364}},
  \bibinfo{pages}{359} (\bibinfo{year}{1985}).

\bibitem[{\citenamefont{Benser et~al.}(1996)\citenamefont{Benser, Marquis, and
  Hudspeth}}]{BMH96}
\bibinfo{author}{\bibfnamefont{M.~E.} \bibnamefont{Benser}},
  \bibinfo{author}{\bibfnamefont{R.~E.} \bibnamefont{Marquis}},
  \bibnamefont{and} \bibinfo{author}{\bibfnamefont{A.~J.}
  \bibnamefont{Hudspeth}}, \bibinfo{journal}{J. Neurosci.}
  \textbf{\bibinfo{volume}{16(18)}}, \bibinfo{pages}{5629}
  (\bibinfo{year}{1996}).

\bibitem[{\citenamefont{Kennedy et~al.}(2005)\citenamefont{Kennedy, Crawford,
  and Fettiplace}}]{KCF05}
\bibinfo{author}{\bibfnamefont{H.~J.} \bibnamefont{Kennedy}},
  \bibinfo{author}{\bibfnamefont{A.~C.} \bibnamefont{Crawford}},
  \bibnamefont{and}
  \bibinfo{author}{\bibfnamefont{R.}~\bibnamefont{Fettiplace}},
  \bibinfo{journal}{Nature} \textbf{\bibinfo{volume}{433}},
  \bibinfo{pages}{880} (\bibinfo{year}{2005}).

\bibitem[{\citenamefont{Kennedy et~al.}(2006)\citenamefont{Kennedy, Evans,
  Crawford, and Fettiplace}}]{KECF06}
\bibinfo{author}{\bibfnamefont{H.~J.} \bibnamefont{Kennedy}},
  \bibinfo{author}{\bibfnamefont{M.~G.} \bibnamefont{Evans}},
  \bibinfo{author}{\bibfnamefont{A.~C.} \bibnamefont{Crawford}},
  \bibnamefont{and}
  \bibinfo{author}{\bibfnamefont{R.}~\bibnamefont{Fettiplace}},
  \bibinfo{journal}{J. Neurosci.} \textbf{\bibinfo{volume}{26(10)}},
  \bibinfo{pages}{2757} (\bibinfo{year}{2006}).

\bibitem[{\citenamefont{K\"{o}ppl and Yates}(1999)}]{KY99}
\bibinfo{author}{\bibfnamefont{C.}~\bibnamefont{K\"{o}ppl}} \bibnamefont{and}
  \bibinfo{author}{\bibfnamefont{G.}~\bibnamefont{Yates}}, \bibinfo{journal}{J.
  Neurosci.} \textbf{\bibinfo{volume}{19(21)}}, \bibinfo{pages}{9674}
  (\bibinfo{year}{1999}).

\bibitem[{\citenamefont{Richter et~al.}(1995)\citenamefont{Richter, Heynert,
  and Klinke}}]{RHK95}
\bibinfo{author}{\bibfnamefont{C.-P.} \bibnamefont{Richter}},
  \bibinfo{author}{\bibfnamefont{S.}~\bibnamefont{Heynert}}, \bibnamefont{and}
  \bibinfo{author}{\bibfnamefont{R.}~\bibnamefont{Klinke}},
  \bibinfo{journal}{Hearing Res.} \textbf{\bibinfo{volume}{83(1-2)}},
  \bibinfo{pages}{19} (\bibinfo{year}{1995}).

\bibitem[{\citenamefont{Cooper and Yates}(1994)}]{CY94}
\bibinfo{author}{\bibfnamefont{N.~P.} \bibnamefont{Cooper}} \bibnamefont{and}
  \bibinfo{author}{\bibfnamefont{G.~K.} \bibnamefont{Yates}},
  \bibinfo{journal}{Hearing Res.} \textbf{\bibinfo{volume}{78 (2)}},
  \bibinfo{pages}{221} (\bibinfo{year}{1994}).

\bibitem[{\citenamefont{Ruggero et~al.}(1997)\citenamefont{Ruggero, Rich,
  Recio, Narayan, and Robles}}]{RRRNR97}
\bibinfo{author}{\bibfnamefont{M.~A.} \bibnamefont{Ruggero}},
  \bibinfo{author}{\bibfnamefont{N.~C.} \bibnamefont{Rich}},
  \bibinfo{author}{\bibfnamefont{A.}~\bibnamefont{Recio}},
  \bibinfo{author}{\bibfnamefont{S.~S.} \bibnamefont{Narayan}},
  \bibnamefont{and} \bibinfo{author}{\bibfnamefont{L.}~\bibnamefont{Robles}},
  \bibinfo{journal}{J. Acoust. Soc. Am.} \textbf{\bibinfo{volume}{101(4)}},
  \bibinfo{pages}{2151} (\bibinfo{year}{1997}).

\bibitem[{\citenamefont{Yates}(1990)}]{Y90}
\bibinfo{author}{\bibfnamefont{G.~K.} \bibnamefont{Yates}},
  \bibinfo{journal}{Hearing Res.} \textbf{\bibinfo{volume}{50(1-2)}},
  \bibinfo{pages}{145} (\bibinfo{year}{1990}).

\bibitem[{\citenamefont{Nadrowski et~al.}(2004)\citenamefont{Nadrowski, Martin,
  and J\"{u}licher}}]{NMJ04}
\bibinfo{author}{\bibfnamefont{B.}~\bibnamefont{Nadrowski}},
  \bibinfo{author}{\bibfnamefont{P.}~\bibnamefont{Martin}}, \bibnamefont{and}
  \bibinfo{author}{\bibfnamefont{F.}~\bibnamefont{J\"{u}licher}},
  \bibinfo{journal}{P. Natl. Acad. Sci.} \textbf{\bibinfo{volume}{101(33)}},
  \bibinfo{pages}{12195} (\bibinfo{year}{2004}).

\bibitem[{\citenamefont{Balakrishnan}(2005)}]{B05}
\bibinfo{author}{\bibfnamefont{J.}~\bibnamefont{Balakrishnan}},
  \bibinfo{journal}{J. Phys. A: Math. Gen.} \textbf{\bibinfo{volume}{38}},
  \bibinfo{pages}{1627} (\bibinfo{year}{2005}).

\bibitem[{\citenamefont{Moreau and Sontag}(2003)}]{MS03}
\bibinfo{author}{\bibfnamefont{L.}~\bibnamefont{Moreau}} \bibnamefont{and}
  \bibinfo{author}{\bibfnamefont{E.}~\bibnamefont{Sontag}},
  \bibinfo{journal}{Phys. Rev. E} \textbf{\bibinfo{volume}{68}},
  \bibinfo{pages}{020901} (\bibinfo{year}{2003}).

\bibitem[{\citenamefont{Vilfan and Duke}(2003)}]{VD03}
\bibinfo{author}{\bibfnamefont{A.}~\bibnamefont{Vilfan}} \bibnamefont{and}
  \bibinfo{author}{\bibfnamefont{T.}~\bibnamefont{Duke}},
  \bibinfo{journal}{Biophys. J.} \textbf{\bibinfo{volume}{85}},
  \bibinfo{pages}{191} (\bibinfo{year}{2003}).

\end{thebibliography}
\end{document}